\documentclass[twocolumn,
                aps,
                superscriptaddress,
                prb, 
                notitlepage,
                10pt,
                longbibliography,
		]{revtex4-2}

\usepackage{amssymb}
\usepackage{amsmath}
\usepackage{dsfont}
\usepackage{graphicx}
\usepackage{dcolumn}
\usepackage{tabularx}
\newcolumntype{C}{>{\centering\arraybackslash}X}
\newcolumntype{L}{>{\raggedright\arraybackslash}X} 
\usepackage{url}
\usepackage[colorlinks=true,breaklinks=true,allcolors=blue]{hyperref}
\usepackage{color}
\usepackage{xcolor}
\usepackage{bbold}
\usepackage[noabbrev]{cleveref}
\Crefname{figure}{Fig.}{Figs.}
\Crefname{equation}{Eq.}{Eqs.}
\Crefname{table}{Tab.}{Tabs.}

\makeatletter
\let\cat@comma@active\@empty
\makeatother

\begin{document}

\title{Leveraging recurrence in neural network wavefunctions for\newline
large-scale simulations of Heisenberg antiferromagnets
on\newline
the square lattice}

\author{M. Schuyler Moss}
\email{msmoss@uwaterloo.ca}
\affiliation{Department of Physics and Astronomy, University of Waterloo, Ontario, N2L 3G1, Canada}
\affiliation{Perimeter Institute for Theoretical Physics, Waterloo, Ontario, N2L 2Y5, Canada}

\author{Roeland Wiersema}
\affiliation{Center for Computational Quantum Physics, Flatiron Institute, 162 Fifth Avenue, New York, NY 10010, USA}

\author{Mohamed Hibat-Allah}
\affiliation{Department of Applied Mathematics, University of Waterloo, Waterloo, ON N2L 3G1, Canada}
\affiliation{Vector Institute, MaRS  Centre,  Toronto,  Ontario,  M5G  1M1,  Canada}

 \author{Juan Carrasquilla } %\orcidlink{0000-0001-7263-3462}
\affiliation{Institute for Theoretical Physics, ETH Zürich, 8093, Switzerland}

\author{Roger G. Melko}
\affiliation{Department of Physics and Astronomy, University of Waterloo, Ontario, N2L 3G1, Canada}
\affiliation{Perimeter Institute for Theoretical Physics, Waterloo, Ontario, N2L 2Y5, Canada}

\begin{abstract}
Machine-learning-based variational Monte Carlo simulations are a promising approach for targeting quantum many-body ground states, especially in two dimensions and in cases where the ground state is known to have a non-trivial sign structure. While many state-of-the-art variational energies have been reached with these methods for finite-size systems, little work has been done to use these results to extract information about the target state in the thermodynamic limit. In this work, we employ recurrent neural networks (RNNs) as a variational ans\"{a}tze, and leverage their recurrent nature to simulate the ground states of progressively larger systems through iterative retraining. This transfer learning technique allows us to simulate spin-$\frac{1}{2}$ systems on lattices with more than 1,000 spins without beginning optimization from scratch for each system size, thus reducing the demands for computational resources. In this study, we focus on the square-lattice antiferromagnetic Heisenberg model, where it is possible to carefully benchmark our results. We show that we are able to systematically improve the accuracy of the results from our simulations by increasing the training time, and obtain results for finite-sized lattices that are in good agreement with the literature values. Furthermore, we use these results to extract accurate estimates of the ground-state properties in the thermodynamic limit. This work demonstrates that RNN wavefunctions are able to extract accurate information about quantum many-body systems in the thermodynamic limit.
\end{abstract}

\maketitle

\section{Introduction}
\label{sec:intro}
The square-lattice spin-$\frac{1}{2}$ antiferromagnetic Heisenberg model (SLAHM) is the prototypical Hamiltonian used to describe quantum antiferromagnets. Besides its relevance to the study of real materials~\cite{PhysRevB.65.144412,PhysRevLett.72.1096,manousakis_spin-_1991}, it is an important proving ground for the development of new computational approaches to the quantum many-body problem.  

Reger and Young were the first to study the ground state of this Hamiltonian with quantum Monte Carlo (QMC) methods, and their seminal work~\cite{reger_monte_1988} provided convincing support of the early results obtained with semi-classical spin-wave theory~\cite{anderson_approximate_1952,kubo_spin-wave_1952,keffer_ferromagnetic_1960}. In particular, QMC simulations confirmed the existence of the proposed but debated long-range antiferromagnetic order of the ground state in the thermodynamic limit.
Over time, the study of the SLAHM ultimately drove the development of other powerful QMC methods such as the stochastic series expansion (SSE) algorithm~\cite{sandvik_quantum_1991,sandvik_finite-size_1997,sandvik_loop_2010,sandvik_stochastic_2019}, which still provides the most accurate estimates of the Hamiltonian's ground-state properties today~\cite{sandvik_finite-size_1997,sandvik_loop_2010,chen_empowering_2024}. 

These QMC simulations provided the understanding that established the SLAHM as a standard benchmark for other computational techniques, such as variational methods like variational Monte Carlo (VMC). 
There have been many VMC studies of the SLAHM~\cite{yokoyama_hubbard_1987,huse_simple_1988,liang_new_1988,liu_variational_1989}, some of which precede the aforementioned QMC results.
However, historically, variational ans\"{a}tze were designed to incorporate specific proposed behaviors of the ground-state wavefunction. These ans\"{a}tze often had very few variational parameters and, as a result, produced results with varying degrees of accuracy. 
In most cases, the proposed ans\"{a}tze captured the long-range antiferromagnetic order, although they typically overestimated the value of the sublattice magnetization in the thermodynamic limit. 
The development of the density matrix renormalization group (DMRG)~\cite{white_density_1992, schollwock_density-matrix_2011, verstraete_density_2023}, allowed for well-understood, stable optimization of more expressive ans\"{a}tze in the form of matrix product states. This approach is efficient and provides state-of-the-art results for ground states of one-dimensional systems, but the computational cost scales unfavorably for two-dimensional systems~\cite{schollwock_density-matrix_2011,verstraete_density_2023}. In order to circumvent the increased demand for resources, DMRG can be performed on square lattices with cylindrical boundary conditions. Such an approach was used to study the SLAHM and the ground-state properties were obtained with a high degree of accuracy~\cite{white_neel_2007}.

In the spirit of increasing the expressiveness of variational ans\"{a}tze, the use of neural networks as parameterized wavefunctions ushered in a new era for variational methods~\cite{carleo_solving_2017}. Capitalizing on the proven ability of neural networks to represent any function~\cite{kolmogorov1957representation,hornik1991approximation,schafer_RNN,le2008representational,zhou2020universality} and modern software and hardware developments, VMC simulations can now be performed using ans\"{a}tze with hundreds of thousands to millions of variational parameters. This progress has led to many impressive results, mostly in the form of the lowest variational energies for finite-sized systems~\cite{roth_high-accuracy_2023,viteritti_transformer_2023,sprague2024,chen_empowering_2024,rende_simple_2024,smith_ground_2024}. Throughout the development of these so called Neural Quantum States (NQS), the SLAHM on a $10\times10$ lattice has been one of the most common benchmarks~\cite{carleo_solving_2017,nomura_restricted_2017,choo_two-dimensional_2019,ferrari_neural_2019,roth_iterative_2020,sharir_deep_2020,kochkov_learning_2021,hibat-allah_supplementing_2024,fu_lattice_2022,chen_systematic_2022,wu_tensor,chen_empowering_2024,wu_variational_2024}. Despite the strong performances of NQS thus far, these variational energies do not provide the full story. 
Indeed, considerations about the properties of ground states in the thermodynamic limit, which are typically a central focus in quantum many-body physics, are rare in the field of NQS.

In this paper, we study the SLAHM with open and periodic boundary conditions. We train powerful autoregressive neural networks known as recurrent neural networks (RNNs) to represent the ground-state wavefunction of this Hamiltonian for lattices with more than 1,000 spins. Using a transfer-learning approach~\cite{zhuang2020transferlearning, zen_transfer_2020} that has been termed ``iterative retraining'' in the context of recurrent neural network wavefunctions~\cite{roth_iterative_2020,hibat-allah_supplementing_2024}, we are able to accurately estimate physical observables from our variational ans\"{a}tze for multiple system sizes without optimizing them from scratch for each system size. This method allows us to overcome prohibitive demands on computational resources and optimization challenges that arise for large system sizes in other neural network architectures.
We perform a finite-size scaling of the ground-state energy and the sublattice magnetization, obtaining extrapolated results that are in close agreement with the values found in the literature. 
Thus we demonstrate that the favorable scaling abilities of RNN wavefunctions make them an advantageous choice of ansatz for studying the ground-state properties of quantum many-body systems in the thermodynamic limit.

\section{Methods}
\label{sec:methods}

We now describe our neural network architecture and optimization strategy. A more in-depth description of the training details, including information on all relevant hyper parameters, is provided in \Cref{app:training_details}. 

\subsection{The Heisenberg antiferromagnet}
\label{methods:slahm}

The SLAHM is defined as, 
\begin{equation}
    \hat{H} = \sum_{\langle i j \rangle} \vec{S}_i\cdot\vec{S}_j,
    \label{hamiltonian}
\end{equation}
where $\langle i j \rangle$ are nearest-neighbor interactions on an $L \times L$ square lattice with $N = L^2$ spin-$\frac{1}{2}$ spins. Importantly, the SLAHM can be made stoquastic using the Marshall-Peierls sign rule~\cite{marshall_antiferromagnetism_1955,capriotti_quantum_2001}, which, by the Perron-Frobenius theorem~\cite{perron_zur_1907, frobenius_uber_1912} renders the ground-state wavefunction sign-free. As such, there is no sign problem for QMC when the Marshall-Peierls sign rule is applied to the SLAHM. Here, we also apply the Marshall-Peierls sign rule so that we can assume a positive variational wavefunction. More details on how the Marshall-Peierls sign rule is imposed can be found in \Cref{app:marshallsign}.

The early predictions about the ground state of the SLAHM obtained from spin-wave theory~\cite{anderson_approximate_1952,kubo_spin-wave_1952} are very accurate, as later confirmed by numerical studies~\cite{reger_monte_1988,sandvik_finite-size_1997}.  
For instance, these calculations first highlighted that quantum fluctuations in the ground state of the SLAHM lower the ground-state energy and reduce (but do not destroy) the sublattice magnetization from the classical value~\cite{anderson_approximate_1952,kubo_spin-wave_1952}. This was the first theoretical prediction that the ground state of the SLAHM has long-range antiferromagnetic order in the thermodynamic limit. The Heisenberg Hamiltonian obeys full SU(2) symmetry, meaning it is invariant under global spin rotation. While any eigenstate of the Hamiltonian will also obey SU(2) symmetry~\cite{marshall_antiferromagnetism_1955,lieb_ordering_1962}, the existence of long-range antiferromagnetic order in the thermodynamic limit implies that this continuous symmetry is spontaneously broken.

In fact, the initial assumption required to apply spin-wave theory to the SLAHM is that the continuous SU(2) symmetry is broken for the ground state in the thermodynamic limit.
In his initial calculations, Anderson crucially (and correctly) assumed a ``large, constant $z$ component of spin on one of the two sublattices''~\cite{anderson_approximate_1952}, thus choosing a direction for the total spin. While this assumption explicitly violates global spin rotation, i.e. the SU(2) symmetry, it can be justified by thinking of the ground state as a superposition, or ``wave-packet'', of SU(2)-symmetric states that yields the desired large, constant $z$ component of the total spin~\cite{anderson_approximate_1952,lhuillier_frustrated_2005,tasaki_long-range_2019}. The energy cost to form such a state decreases as $1/N$, meaning this assumption and subsequent interpretation is valid in the thermodynamic limit where $N\rightarrow\infty$. In other words, there exists a set of states $\vert 0 \rangle$, which includes the ground state and some low-lying excited states that collapse onto the ground state in the thermodynamic limit as $1/N$~\cite{anderson_approximate_1952,hasenfratz_finite_1993}. This set of states $\vert 0 \rangle$ is known as Anderson's tower of states and describes the spontaneous SU(2) symmetry breaking of the SLAHM. Interestingly, from the symmetry-broken ground state, spin-wave theory correctly predicted the emergence of gapless magnon excitations. These excitations were later identified as the Goldstone modes associated with the continuous symmetry breaking exhibited by the SLAHM~\cite{goldstone_broken_1962}.

\subsection{Variational optimization}
\label{methods:vmc}

Variational Monte Carlo (VMC) simulations of quantum many-body systems involve defining an appropriate variational ansatz $\vert\Psi_\mathcal{W}\rangle$ with trainable parameters $\mathcal{W}$~\cite{becca_quantum_2017}.
For a given Hamiltonian $\hat{H}$, we can calculate a variational energy as
\begin{align}
    E_\mathcal{W} &\equiv \frac{\langle \Psi_\mathcal{W}\vert \hat{H}\vert \Psi_\mathcal{W}\rangle}{\langle \Psi_\mathcal{W}\vert \Psi_\mathcal{W}\rangle}\nonumber
    \\
    & \approx \frac{1}{N_s}\sum_{\vec{\sigma}\sim p_\mathcal{W}(\vec{\sigma})} H_\text{loc}(\vec{\sigma}),
    \label{eq:variationalenergy}
\end{align}
which involves importance sampling $N_s$ spin configurations $\vec{\sigma}\in\{0,1\}^N$ from $p_\mathcal{W} \equiv |\psi_\mathcal{W}|^2$. Here we have introduced the local energy,
\begin{align}
    H_\text{loc}(\vec{\sigma}) \equiv \frac{\langle \vec{\sigma}\vert\hat{H}\vert\Psi_\mathcal{W}\rangle}{\langle \vec{\sigma}\vert\Psi_\mathcal{W}\rangle}.
    \label{eq:localenergies}
\end{align}
As a consequence of the variational principle, the variational energy $E_\mathcal{W}$ provides an upper bound to the true ground-state energy. We can thus approximate the true ground state of $\hat{H}$ with our variational ansatz $\vert\Psi_\mathcal{W}\rangle$ by minimizing the variational energy with an appropriately chosen optimization scheme~\cite{becca_quantum_2017,hibat-allah_recurrent_2020}. 
In this work, we use the Adam optimizer~\cite{kingma2017adammethodstochasticoptimization} to minimize $E_\mathcal{W}$.

The estimation of the variational energy occurs at every step of the optimization. Therefore, the ease with which one can compute the local energies plays an important role in a VMC simulation. The cost of computing \Cref{eq:localenergies} scales with the number of off-diagonal terms in the Hamiltonian because the variational ansatz must be evaluated for each sample $\vec{\sigma}^{\prime}$ connected to the original sample $\vec{\sigma}$ by $\hat{H}$~\cite{schmitt2020quantum}. The cost associated with each evaluation of the variational wavefunction will be discussed in the following sections. For the SLAHM the number of off-diagonal terms is equal to the number of nearest-neighbor interactions, which scales as $\mathcal{O}(N)$. 

\subsection{Recurrent neural network wavefunctions}
\label{methods:rnns}

RNNs are a powerful class of autoregressive neural networks that were proposed as variational ans\"{a}tze for ground states due to their success in modeling long sequences of correlated variables~\cite{hibat-allah_recurrent_2020,sprague2024}. 
Instead of  directly learning the joint probability distribution over a full spin configuration $ \vec{\sigma}= (\sigma_1,\dots,\sigma_N)$, autoregressive neural networks make use of the chain rule of probabilities to instead learn conditional probabilities over the individual spins $\sigma_i$,
\begin{equation}
    p(\vec{\sigma}) 
    = p(\sigma_1)p(\sigma_2\vert\sigma_1)\dots p(\sigma_N\vert\sigma_{N-1},\dots,\sigma_2,\sigma_1)\,.
    \label{chain_rule_of_prob}
\end{equation}
Recasting the learning task in this way enforces the overall normalization of the full joint probability distribution. This allows for efficient sampling: independent samples for each spin $\sigma_i$ can be obtained directly from the corresponding conditional probability $p(\sigma_i\vert\sigma_{j<i})$. The ability to obtain independent samples can alleviate the sampling costs associated with computing the variational energy in \Cref{eq:variationalenergy}.

\begin{figure}
    \centering
    \includegraphics[width=0.35\textwidth]{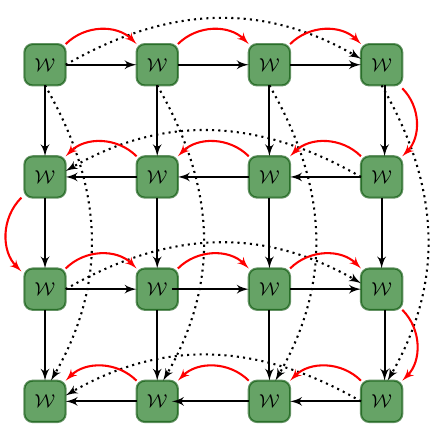}
    \caption{\small A 2D RNN wavefunction defined for a square lattice with $L=4$. The autoregressive sequence is defined by the red arrows. Sampling and inference are performed along this path. The information in the network, stored in the hidden vectors, is passed in two directions, along the black arrows. The black dotted arrows show how pseudo-periodic boundary connections can be built in to the RNN wavefunction. Both the two-dimensional information passing and the pseudo-periodic boundary connections are implemented in a causal way such that the autoregressive sequence is not violated.}
    \label{fig:RNN}
\end{figure}

While any neural network can be made autoregressive using masking~\cite{germain2015mademaskedautoencoderdistribution,uria2016neuralautoregressivedistributionestimation,bortone_impact_2024}, RNNs have another desirable feature, which is their \emph{recurrent} nature. 
The elementary building block of an RNN is a \emph{recurrent cell}, depicted by the green boxes in \Cref{fig:RNN}. These cells represent a non-linear transformation which involve the tuneable parameters of the model $\mathcal{W}$ and act on the input hidden vectors $\vec{h}_{\text{input}}$ and local spins $\sigma_{\text{input}}$. The size of the hidden vectors $d_h$ is a choice and controls the expressiveness of the network, as these vectors are the conduit through which information is passed in the RNN. The output of the cell is a new hidden vector $\vec{h}_{\mathrm{out}}$ of length $d_h$, which is then used to obtain the probability distribution over the corresponding spin $p_\mathcal{W}(\sigma_i)$.  This process is repeated for all $i$, using the same cell and the same set of weights $\mathcal{W}$. The recurrent nature of RNNs thus allows us to define a variational ansatz with a total number of parameters that is independent of the system size in the sense that the shapes of the weight matrices do not explicitly depend on the number of spins in the system $N$. We emphasize that the choice of $d_h$ alone controls the number of parameters in the RNN wavefunction when weights are shared as described.

\Cref{fig:RNN} shows the basic structure of a two-dimensional (2D) RNN wavefunction. 
In order to construct a wavefunction from an RNN, one must obtain the probability distributions over each spin configuration. \Cref{chain_rule_of_prob} requires that this computation is performed sequentially, meaning one must choose a one-dimensional (1D) path along which samples are obtained or inference is performed. This 1D path, or the so-called \emph{autoregressive sequence}, is highlighted in red. The autoregressive sequence is a choice and can play an important role in the convergence of the training~\cite{teoh_autoregressive_2024}. Autoregressive sampling and inference scale as $\mathcal{O}(Nd_h^2)$. For the entirety of this work, we fix $d_h=256$ so that the factor of $d_h^2$ can be treated as a constant and the main scaling comes from $N$ alone. In the NQS setting, performing inference means evaluating the variational wavefunction for a given sample, i.e. obtaining $\psi_{\mathcal{W}}(\vec{\sigma}) = \langle \vec{\sigma}\vert\psi_{\mathcal{W}}\rangle$.

The black arrows in \Cref{fig:RNN} show how the hidden vectors $\vec{h}_{\mathrm{out}}$ and the spins $\sigma_i$ are passed between RNN cells. 
The RNN was originally introduced to model 1D sequences~\cite{lipton_critical_2015} and information was passed only along the autoregressive path.
For 2D physical systems, it is desirable to use the 2D generalization, where hidden vectors and local variables can be passed in more than one direction while abiding by the autoregressive sequence~\cite{hibat-allah_recurrent_2020}. When using an RNN wavefunction to represent the ground state of a system with periodic boundary conditions, it is also useful to build pseudo-periodic connections into the architecture~\cite{luo_gauge-invariant_2023,hibat-allah_investigating_2023,hibat-allah_recurrent_2024}. Both of these generalizations, and the extra recurrent connections they introduce, incorporate important correlations between physical degrees of freedom into the ansatz.

After obtaining all of the conditional probability distributions over each spin according to the predefined autoregressive sequence, one can construct the variational wavefunction as follows: when a ground state has real and positive amplitudes, as is the case for the SLAHM when the Marshall-Peierls sign rule is applied, the conditional amplitude can be obtained directly from the conditional probability, i.e. $\psi(\sigma_i|\sigma_{j<i}) = \sqrt{p_{\mathcal{W}}(\sigma_i|\sigma_{j<i})}$. The amplitude of the full spin configuration $\vec{\sigma}$ is the product of these local amplitudes $\psi(\vec{\sigma}) = \prod_i \psi(\sigma_i|\sigma_{j<i})$. \Cref{app:rnn_details} contains a more detailed description of the RNN wavefunctions used in this work.

\subsection{Symmetries}
\label{methods:symmetries}

Incorporating symmetries in a variational ansatz can dramatically reduce the size of the effective Hilbert space and improve variational accuracies~\cite{choo_symmetries_2018, nomura_helping_2021}. For the SLAHM, two types of symmetries can be exploited. The first are those that arise due to the Hamiltonian. Efficiently imposing the full SU(2) symmetry of the SLAHM and its eigenstates in NQS is possible by working in the angular momentum basis~\cite{vieijra_restricted_2020,luo_gauge-invariant_2023}.
However, it is known that the finite-size ground states of the SLAHM are singlet states, which also obey U(1) symmetry~\cite{marshall_antiferromagnetism_1955,lieb_ordering_1962}. U(1) symmetry can be explicitly enforced in the $z$ basis by constraining the state to be in the correct magnetization sector~\cite{hibat-allah_recurrent_2020, morawetz_u1-symmetric_2021, hibat-allah_supplementing_2024}. For simplicity, we adopt the latter approach.

For certain lattices, point-group symmetries can also be exploited to improve variational calculations~\cite{nomura_helping_2021, hibat-allah_supplementing_2024}. 
While it is possible to exactly enforce these symmetries such that the variational wavefunction is restricted to the correct symmetry sector, this approach has only been studied for certain architectures such as restricted Boltzmann machines~\cite{choo_symmetries_2018}, convolutional neural networks~\cite{choo_two-dimensional_2019}, or group convolutional neural networks~\cite{roth_group_2021, roth_high-accuracy_2023}. 
Strictly enforcing point-group symmetries using RNN wavefunctions would either violate the autoregressive property or be computationally intractable; however, we can still train the wavefunction to learn invariance under a given symmetry group during the training with a ``data-augmentation'' approach~\cite{Reisert2007equiv,Yarotsky2022symmetrization}. This approach involves averaging the amplitudes over the action of the group $\mathcal{G}$. More specifically, 
we can obtain the symmetrized amplitude $\psi_\mathcal{W}^\prime(\vec{\sigma}) = \sqrt{p_\mathcal{W}^\prime(\vec{\sigma})}$, where
\begin{align}
     p_\mathcal{W}^\prime(\vec{\sigma}) = \frac{1}{|\mathcal{G}|} \sum_{\mathcal{T} \in \mathcal{G}} ^{|\mathcal{G}|}p_\mathcal{W}(\mathcal{T}\vec{\sigma}).
     \label{eq:sym}
\end{align}
Here, $\vert\mathcal{G}\vert$ is the number of elements in the group $\mathcal{G}$ and  $\mathcal{T}\vec{\sigma}$ is a symmetry-transformed spin configuration. 
Note that this symmetrization scheme also applies to quantum states with non-trivial sign structure~\cite{hibat-allah_recurrent_2020, sharir_deep_2020} and the symmetrized ansatz is compatible with perfect sampling~\cite{Reh_optimizing_2023}.  While it has been shown that the inclusion of lattice symmetry improves the convergence of variational optimization, this method increases the computational cost by a factor of $\vert\mathcal{G}\vert$ due to the additional evaluations of the variational ansatze needed to compute $p^\prime_\mathcal{W}(\vec{\sigma})$. For the square lattice, with both open and periodic boundary conditions, we enforce $C_{4v}$ point-group symmetry using \Cref{eq:sym}.

\subsection{Iterative retraining}
\label{methods:retraining}

Understanding the ground states of physical systems in the thermodynamic limit is one of the central tasks of quantum many-body physics. Computationally, this is achieved by performing finite-size scaling studies, which involve obtaining estimates of ground-state properties for many different system sizes. For machine-learning-based VMC, obtaining results for different system sizes typically means training an NQS from scratch for each system size, which can be extremely demanding in terms of computational resources. One of the promises of modern machine learning is the ability of neural networks to \emph{generalize} beyond the data it is trained on. In the context of NQS, this might translate to training a neural network to represent a ground state for a system of one size, and then using the same network to study the a ground state for the same system but of another size. The ability for NQS to generalize in this way would make it more efficient to obtain the results necessary for finite-size scaling studies. 

As mentioned, the recurrent nature of RNN wavefunctions can be leveraged so that the number of trainable parameters is independent of the size of the physical system. 
This property allows the optimal weights from a VMC simulation performed for one system size to be used as the initial weights for a VMC simulation for a larger system size.
In practice, this reuse of weights provides a good starting point for the variational optimization for the larger system size, demonstrating that these ans\"{a}tze are indeed able to generalize well to larger system sizes. 
Roth et. al.~\cite{roth_iterative_2020} was the first to propose this ``iterative retraining'' technique, showing favorable generalization behavior for RNN wavefunctions. More specifically, they found that the variational optimization for larger system sizes converged with a decreasing number of training steps when reusing weights in this way, thus reducing demands for computational resources. In the machine learning community, this technique is considered a form of \emph{transfer learning} because the model learns something during one training task (a VMC simulation for a given system size) that is transferable and useful to another training task (a VMC simulation for a larger system size). 

Convolutional neural networks and graph neural networks also have a number of parameters that is independent of system size and could be iteratively retrained~\cite{choo_two-dimensional_2019,yang_scalable_2020,kochkov_learning_2021}. To our knowledge, this transfer learning method has only been studied for graph neural networks, showing similar successes as with RNN wavefunctions~\cite{yang_scalable_2020,kochkov_learning_2021}. While transfer learning is possible for network architectures that are system-size-dependent, one must perform non-trivial manipulations of the weights which exhibit varying degrees of success~\cite{zen_transfer_2020}. 

It is worth mentioning that even though the number of parameters in an RNN wavefunction with shared weights does not explicitly depend on the number of spins in the system $N$, the hidden dimension $d_h$ should display some scaling with $N$ in order to reach sufficiently accurate results. For iterative retraining to be successful, one must choose a sufficiently large $d_h$, so that the RNN wavefunction is expressive enough to represent the ground state for the largest system considered. The exact scaling of $d_h(N)$ is an open question, though it has been studied for other architectures~\cite{dan_scaling_rbm}.

\begin{figure}
    \centering
\includegraphics[width=\linewidth]{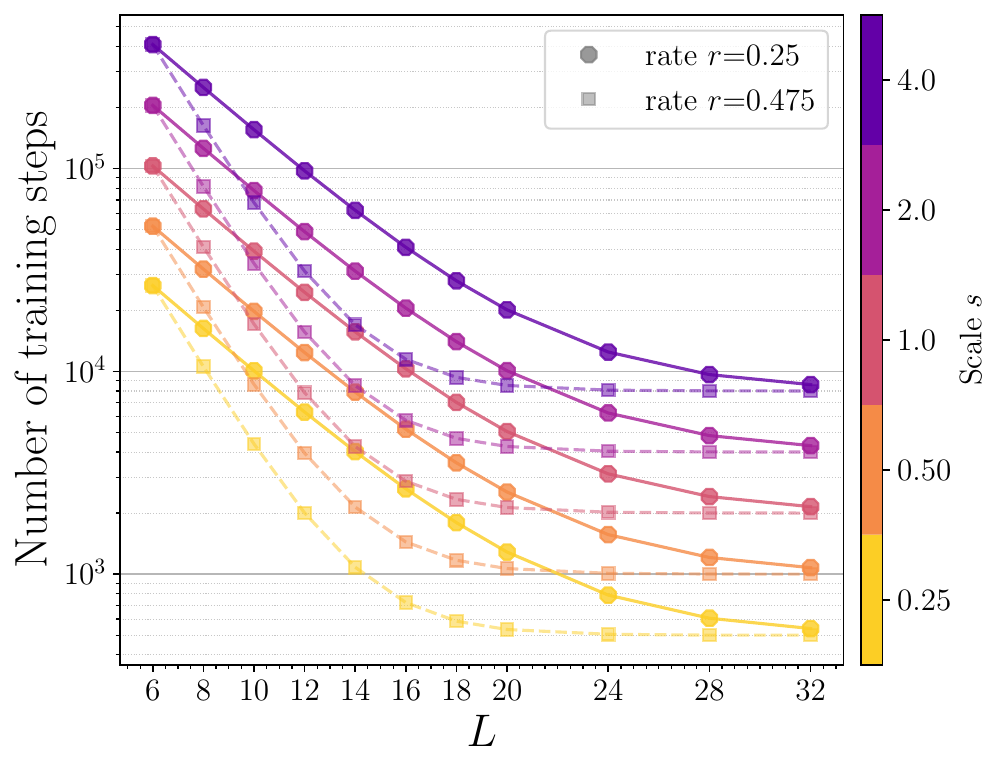}
    \caption{The number of training steps used in the optimization for each system size as determined by our parameterized training schedule defined in \Cref{eq:schedule}. We consider five different scales and two rates. The colors and markers are used to indicate these two parameters respectively. Throughout this work when results from all of our simulations are presented, this legend will be used. Note that when $s=1$ and $r=0.475$, our training schedule is similar to that which has been used in other iterative retraining studies~\cite{roth_iterative_2020,hibat-allah_supplementing_2024}.}
    \label{fig:steps_per_L}
\end{figure}

We employ iterative retraining for our RNN wavefunctions and emphasize that this method allows us to obtain ground-state estimates for very large system sizes at a significantly lower computational cost.
As we retrain our RNN for larger lattices, we decrease the number of training steps according to a parameterized exponential decay which was inspired by the training schedules used in previous works~\cite{roth_iterative_2020,hibat-allah_supplementing_2024}. We define the number of steps for a given system size $L$ as
\begin{align}
\label{eq:schedule}
    N_\text{steps} (L, s, r; L_0, C, F) = s \times \big[ C\, \text{exp}(-r(L &-L_0))+ F \big],
\end{align}
where $s$ is an overall \emph{scale} factor, $r$ is the \emph{rate} at which the number of steps per $L$ decays.  We fix $C$, which adjusts the number training steps for the smallest system size $L_0$, and $F$, which determines the number of training steps for very large lattice sizes. The smallest system size we consider is $L_0=6$. \Cref{fig:steps_per_L} shows the training schedules considered in this work. All hyperparameters, including values for $C$, and $F$ are reported in \Cref{app:training_details}. Importantly, the parameterized schedule of \Cref{eq:schedule} gives us a way to directly control the number of training steps, and therefore the total training time, of our simulations. For more details about how this training schedule translates to the overall run time of our simulations, see \Cref{app:runtimes}.

\section{Results and discussion}
\label{sec:results}

Combining all of the methods described above, we estimate ground-state properties in the thermodynamic limit by performing a finite-size scaling study. 
In particular, we examine the variational energies and estimates of the sublattice magnetization obtained from our trained ans\"{a}tze for system sizes up to $L=32$, where there are $N>1,000$ spins.

We focus on our results for the SLAHM with periodic boundary conditions only, as it is the case which has been studied more thoroughly both theoretically and numerically. We compare against reference values from various QMC studies. The reference ground-state energies for finite-size systems will appear in a forthcoming manuscript~\cite{anders_unpublished}. The reference value for the ground-state energy in the thermodynamic limit is from Ref.~\cite{sandvik_finite-size_1997}. All estimates for the squared sublattice magnetization are taken from Ref.~\cite{sandvik_loop_2010}. We take these reference values to be ``ground truth''. We performed a similar analysis for the SLAHM with open boundary conditions, which both supports and expands on previous RNN results~\cite{roth_iterative_2020,hibat-allah_supplementing_2024}. Those results can be found in \Cref{app:OBC}.

\subsection{Energy scaling}
\label{results:energy}

The first quantity to assess when using a variational method is the variational energy. When the energy error is much smaller than the energy gap between the ground state and the first excited state, one can safely assume that the largest contribution to the variational wavefunction is the overlap with the true ground-state wavefunction ~\cite{becca_quantum_2017}. Within this regime, the accuracy of observables can be also estimated~\cite{becca_quantum_2017,beach_making_2019}.

It is known that the first excited state in Anderson's tower of states $\vert 0\rangle$ for the SLAHM on a finite-size lattice is a triplet state~\cite{marshall_antiferromagnetism_1955} and that the energy gap between the singlet ground state and this triplet excited state closes as $1/N$~\cite{hasenfratz_finite_1993}.
Although precise values of the energy gaps for finite-size systems are not available, the known scaling behavior informs us that we must have increasingly accurate variational energies to guarantee convergence close to the ground state.

\begin{figure}
   \includegraphics[width=\linewidth]{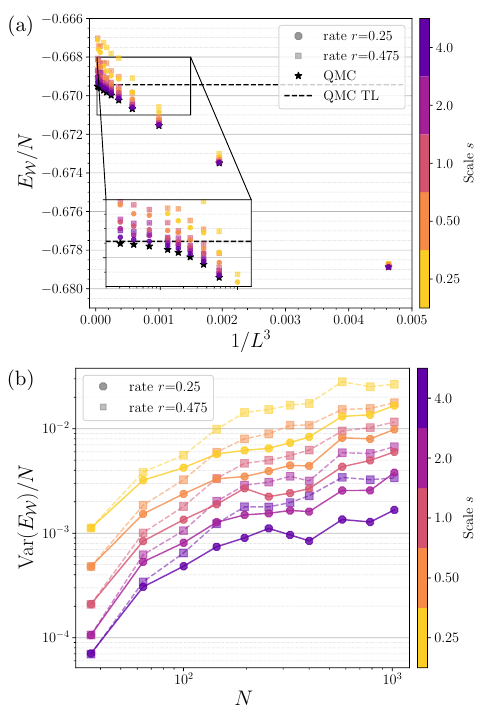}
    
    \caption{\small 
    (a) The final variational energies obtained from all optimized RNN wavefunctions for each system size plotted according to the known scaling form of the energy, $1/L^3$~\cite{neuberger_finite-size_1989,hasenfratz_finite_1993}. All of the final variational energies were estimated with $10\times10^3$ samples. Reference energies for each $L$~\cite{anders_unpublished} and the thermodynamic limit (TL) value of the ground-state energy per spin~\cite{sandvik_finite-size_1997} from QMC is shown for reference. The inset shows a zoomed view of the region close to zero plotted on a log scale for easier viewing.  
    (b) The variances of the above final variational energies as a function of the number of spins in the system $N$.
    }
    \label{fig:energies_peri}
\end{figure}

\Cref{fig:energies_peri}(a) shows the variational energies obtained from all of our simulations compared against the available reference values. 
As the lattice size increases, the variational energies from some simulations appear to diverge from the QMC reference values. However, we find that the variational energies and their scaling behavior can be improved systematically using the two parameters that control the training schedule defined in \Cref{eq:schedule}. Specifically, as we increase the scale $s$ or decrease the decay rate $r$, both of which increase the overall training time, the variational energies tend towards the correct values and show better scaling behavior. 

To further demonstrate the systematic improvement of our results, we also examine the variance of the variational energies, which is a way to gauge how close a variational wavefunction is to an eigenstate of the Hamiltonian~\cite{Gross_criterion_1990, Assaraf_zerovar_2003, wu_variational_2024}. If the variational wavefunction has perfectly represented the ground state, then the measured variational energy will be exactly the ground-state energy with zero variance. \Cref{fig:energies_peri}(b) shows the variances of the variational energies obtained from all individual simulations. Again, we observe that $s$ and $r$ can be tuned to improve the results. Longer training time leads to systematically lower variances of the variational energy. 

Recently, Wu \emph{et al.} introduced another metric, the V-score, which takes into account both the variational energy and the variance of the variational energy~\cite{wu_variational_2024}. We examine our V-scores in \Cref{app:vscore}.

\begin{figure}
    \includegraphics[width=\linewidth]{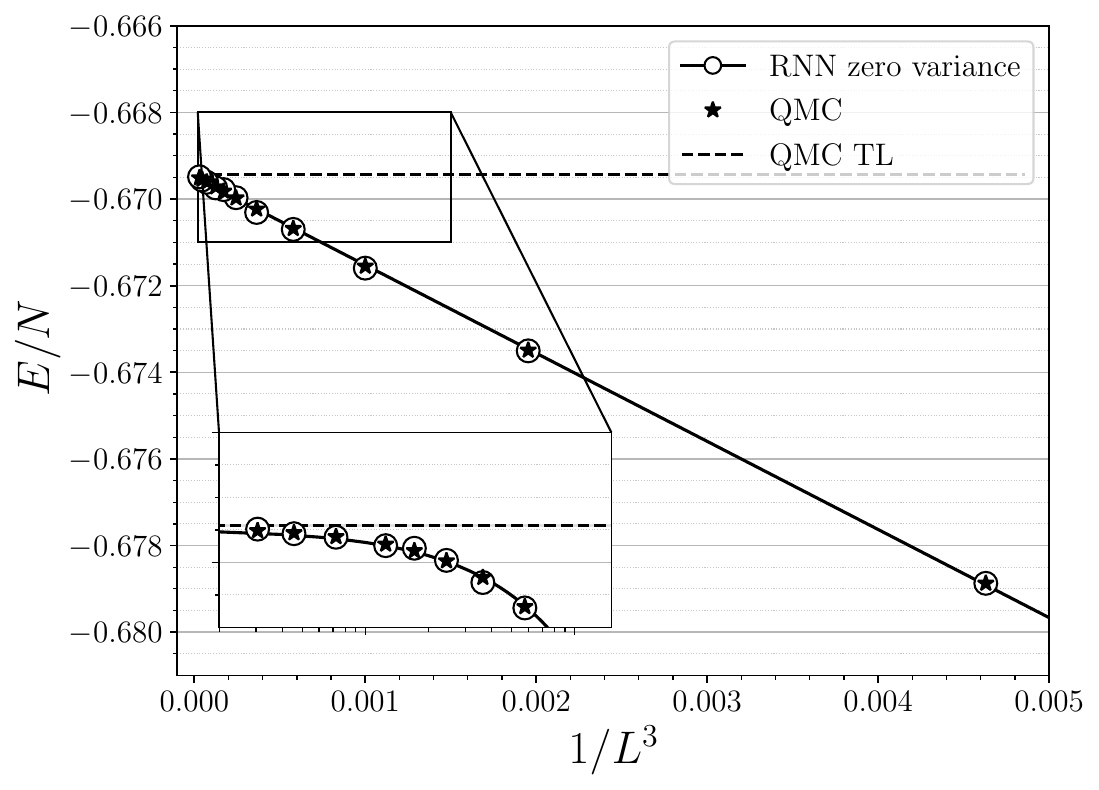}
    \caption{\small Energy estimates obtained from extrapolating the variational energies to their zero-variance value compared to reference energies obtained from QMC simulations~\cite{anders_unpublished}. The thermodynamic limit (TL) value of the ground-state energy per spin from QMC~\cite{sandvik_finite-size_1997} is shown for reference. The inset shows a zoomed view the region close to zero plotted on a log scale for easier viewing. 
    }
    \label{fig:zero_var_energy}
\end{figure}

\Cref{fig:energies_peri}(a) and (b) show that our results are made systematically better by either increasing the scale $s$ or decreasing the rate $r$ of the training schedule. When one has such a convergent sequence of states, it is possible to extrapolate the variational energy to the zero-variance limit~\cite{kwon_effects_1998,becca2000stabilitydwavesuperconductivitytj,Sorella2001,hu_direct_2013,nomura_restricted_2017,fu_variance_2023}. To obtain estimates of zero-variance energies and appropriate error bars from our variational energies, we first perform a linear fit through the best of the variational energies and their respective variances for a given lattice size $L$~\footnote{We extrapolated our results from simulations with the three largest scales $s=1.0,2.0,4.0$, which produced the most accurate variational energies, as seen in the inset of \Cref{fig:energies_peri}(a), as well as the smallest variances on the variational energies, as seen in \Cref{fig:energies_peri}(b).}. See \Cref{app:zero-var} for an example of our zero-variance extrapolation for $L=32$. Using that linear fit and a wild bootstrap method~\cite{wu_bootstrap}, with $10^3$ bootstraps, we obtain more statistically accurate estimates for the zero-variance energies and their error bars.
The resulting zero-variance energies, shown in \Cref{fig:zero_var_energy}, are in close agreement with the reference energies from QMC. 

Next, we perform a finite-size scaling of the ground-state energies using the zero-variance energy estimates for each lattice size. 
For the SLAHM with periodic boundary conditions, the finite-size behavior of the ground-state energy has the following form ~\cite{neuberger_finite-size_1989, hasenfratz_finite_1993}:  
\begin{align}
    E(L) = E_{\infty} + \frac{e_1}{L^3} + \mathcal{O}\left(\frac{1}{L^4} \right),\label{eq:El}
\end{align}
which has been confirmed by rigorous numerical studies~\cite{sandvik_finite-size_1997}. 
We fit \Cref{eq:El} to our zero-variance energies for all values of $L$ using a linear regression. We again improve our statistical results with the wild bootstrap method~\cite{wu_bootstrap} and $10^3$ bootstraps.  
We note that our extrapolated value can be improved by excluding energies from the smaller values of $L$ when fitting \Cref{eq:El}. This latter approach accounts for finite-size effects, which are more significant for smaller lattices. However, the results of our simulations are most accurate for these smaller lattices, so there is a delicate trade-off between prioritizing larger system sizes or prioritizing high-accuracy contributions to the fit. As such, we perform our extrapolation using all available values of $L$. 
Our extrapolated value for the ground-state energy is $E_\infty=-0.6694886(5)$, which has a relative error of $5\times10^{-5}$ with respect to the reference QMC value $E_\infty^{\text{QMC}}=-0.669437(5)$~\cite{sandvik_finite-size_1997}.

\subsection{Sublattice magnetization scaling}
\label{results:magnetization}

Although the variational energy gives insight into the accuracy of a simulation, it does not characterize any property of the ground state itself. On the other hand, correlation functions can be used to understand collective behavior in a system, such as the phase to which a ground state belongs. In this work, we estimate the real-space correlations of the SLAHM using our optimized wavefunctions. The two real space correlations we measure are
\begin{align}
    C(i,j) = \left\langle \vec{S}_i\cdot\vec{S}_j \right\rangle,
    \label{eq:real_space_corr}\\
    C^z(i,j) = 3\left\langle S^{z}_i S^{z}_j \right\rangle.
    \label{eq:real_space_corr_z}
\end{align}
Since the Heisenberg model and its finite-size ground states obey SU(2) symmetry, estimates based on \Cref{eq:real_space_corr} and \Cref{eq:real_space_corr_z} should be equal. $S^z$ operators are diagonal in the computational basis, making $C^z(i,j)$ correlations much less costly to compute. As such, it is common to exploit the rotational invariance and calculate only the $C^z(i,j)$ correlations. 

Using the correlations defined in \Cref{eq:real_space_corr}, we can compute the correlations in momentum space as follows:
\begin{align}
    S^{(z)}(L,\vec{q}) = \frac{1}{L^2} \sum_{i,j} e^{i\vec{q}\cdot\vec{r}} C^{(z)}(i,j),
    \label{eq:structure_factor}
\end{align}
where $\vec{r} = \vert \vec{r}_i - \vec{r}_j\vert$. The superscript $(z)$ indicates which real space correlations defined in \Cref{eq:real_space_corr} or \Cref{eq:real_space_corr_z} were used. Of particular interest is the staggered structure factor, which is defined as $S^{(z)}\left(L,\vec{q}=(\pi,\pi)\right)$.

Ultimately, we are interested in the sublattice magnetization of the SLAHM in the thermodynamic limit. The sublattice magnetization is the antiferromagnetic order parameter in the ground state of the SLAHM. In particular, it is typical to measure the squared sublattice magnetization $M^2$~\cite{reger_monte_1988,sandvik_finite-size_1997}.
For the SLAHM with periodic boundary conditions, this quantity can be estimated in two ways. The first is based on the staggered structure factor defined by \Cref{eq:structure_factor},
\begin{align}
    M^2_{(z)}(L) = \frac{S^{(z)}(L,\vec{q} = (\pi,\pi))}{L^2}.
    \label{eq:M}
\end{align}
Alternatively, one can estimate the same quantity using only the real space correlations between spins with the longest separation vector,
\begin{align}
    M^2_{C^{(z)}}(L) = C^{(z)}(L/2,L/2).
    \label{eq:M_C}
\end{align}

\begin{figure}
    \includegraphics[width=\linewidth]{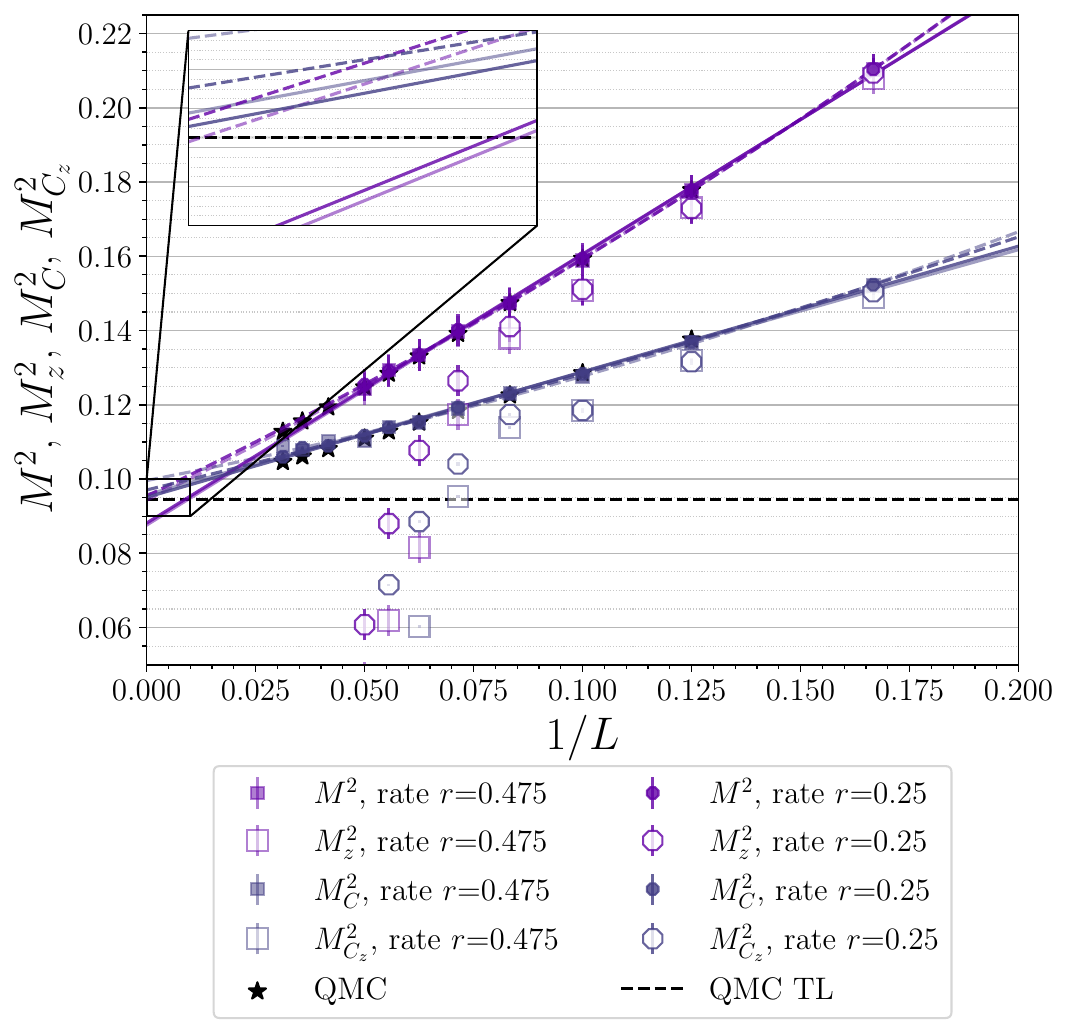}
    \caption{\small
    Estimates of the squared sublattice magnetization scaled as a function of $1/L$. Estimates of $M^2$ and $M_C^2$ are shown to demonstrate that they should extrapolate to the same value in the thermodynamic limit. The inset shows a zoomed view of the $y$ intercept. Closed markers represent the values of $M^2$ and $M_C^2$ estimated using the correlations defined by \Cref{eq:real_space_corr}, while open markers represent the values of $M^2$ and $M^2_C$ that are estimated using the correlations defined by \Cref{eq:real_space_corr_z}. Finite-size estimates and the thermodynamic limit (TL) value of the squared sublattice magnetization from QMC~\cite{sandvik_loop_2010} are shown for reference.
    }
    \label{fig:magnetization}
\end{figure}

\Cref{fig:magnetization} shows the estimates of \Cref{eq:M} and \Cref{eq:M_C} obtained from our most accurate simulations (scale $s=4.0$). Each real-space correlation was estimated with $10\times10^3$ samples.
We only calculate $M^2$ for $L\leq20$ because the calculation scales as $\mathcal{O}(N^3)$: \Cref{eq:structure_factor} involves a sum over all two-point correlations and performing inference with our RNN wavefunction scales as $\mathcal{O}(N)$.

Our finite-size estimates of $M^2$ and $M^2_C$ appear to be in close agreement to the available QMC reference values up to very large system size $L$; however, our estimates of $M^2_z$ and $M_{C^z}^2$ do not exhibit the expected behavior. The growing disagreement between $M^2$ and $M^2_z$ (or equivalently between $M^2_C$ and $M^2_{C^z}$) indicates that the SU(2) symmetry of our RNN wavefunction is broken during the iterative retraining. 
The symmetry breaking behavior is striking for both rates shown in \Cref{fig:magnetization}, but we notice that the divergence between e.g. $M^2$ and $M^2_z$ for $r=0.25$ is slower than the divergence for $r=0.475$. In general, we notice that for larger scales or slower rates (smaller r), the symmetry breaking occurs later and less dramatically. It is possible that even longer training time might allow our RNN wavefunctions to restore the broken SU(2) symmetry.
Interestingly, we only observe this symmetry breaking in our simulations for systems with periodic boundary conditions (see \Cref{fig:magnetization_open} in \Cref{app:OBC}). It is possible that this behavior is the result of the expressiveness (or lack thereof) of the recurrent cell employed for the periodic RNN wavefunctions (see \Cref{app:rnn_details}). To this point, some studies have shown that broken symmetries in an NQS can be restored by increasing the expressiveness of the ansatz~\cite{ferrari_neural_2019}. 

Despite the observed symmetry breaking, we are still able to extract accurate estimates of the sublattice magnetization in the thermodynamic limit from the estimates of $M^2$ and $M^2_C$ from our simulations. To obtain our extrapolated values of the sublattice magnetization, we perform multiple polynomial fits including up to the third order in $1/L$.
Sandvik et. al~\cite{sandvik_finite-size_1997} performed a constrained third-order polynomial regression and argued that the additional sub-leading terms were essential for obtaining an accurate estimate of the extrapolated value. 
In ~\Cref{fig:magnetization}, we show only the linear and second-order polynomial fits because we find that fitting a third-order polynomial to our data leads to spurious behavior outside of the data range.

Fitting a second-order polynomial to $M^2$ yields extrapolated values for $M_\infty$ that are within the error bars of $M^C_\infty$, the values obtained from performing either a first- or second-order fit on $M^2_c$. The agreement between these extrapolated values can be seen in the inset of \Cref{fig:magnetization}. The linear fit on $M^2_c$ provides tighter error bars, so we report $M^C_\infty$ obtained from that fit. Our extrapolated estimates for $M^2$ and $M^2_C$ are not only in good agreement with each other, but also with the reference value obtained from QMC simulations $M_\infty^{\text{QMC}} = 0.30743(1)$~\cite{sandvik_loop_2010}. We obtain extrapolated values for the sublattice magnetization of $M_\infty = 0.309(2)$ and $M_\infty^C = 0.308(2)$ using a second-order polynomial fit on our estimates of $M^2$ and a linear fit on our estimates of $M^2_C$ and then improving our statistical estimates with a wild bootstrap~\cite{wu_bootstrap} method and $10\times10^3$ bootstraps.

Our results for the sublattice magnetization strongly suggest that the states learned by our RNN wavefunctions are a superposition of the states in Anderson's tower of states $\vert 0\rangle$.
Anderson's tower of states was an idea originally proposed to offer a mechanism that explains the spontaneous SU(2) symmetry breaking in the SLAHM~\cite{anderson_approximate_1952,tasaki_long-range_2019}. 
Importantly, all of the lowest-lying states in $\vert0\rangle$ exhibit the same squared value of the sublattice magnetization, i.e. the same values for $M^2$, $M^2_{z}$, $M^2_C$, and $M^2_{C^{z}}$~\cite{lhuillier_frustrated_2005,bernu_signature_1992,tasaki_long-range_2019}.
The observed symmetry breaking in our RNN wavefunctions means that our wavefunctions contain contributions from states other than the true finite-size ground state, which is an SU(2)-symmetric singlet state. This is consistent with the fact that the variances of our variational energies are non-zero. However, our ability to extract accurate estimates of $M_\infty$ and $M_\infty^C$ suggests that the extra contributions are dominated by the states in $\vert0\rangle$.

\subsection{Do better variational energies guarantee better correlations?}
\label{results:errors}

Thus far, we have demonstrated that the results of our variational simulations are accurate and can be systematically improved. We have shown that the accuracy of both the energies and the correlations captured by our ans\"{a}tze improve with longer training times, controlled by the scale $s$ and rate $r$ in our training schedule defined by~\Cref{eq:schedule}. Because both of these estimates exhibit the same improvement as a result of adjusting the training schedule, it might seem reasonable to assume that the accuracies of the two estimates are correlated. However,  because we do not know the precise values of the gap above the ground-state energies, this assumption may not be true and should therefore be examined directly.
\begin{figure}
    \includegraphics[width=\linewidth]{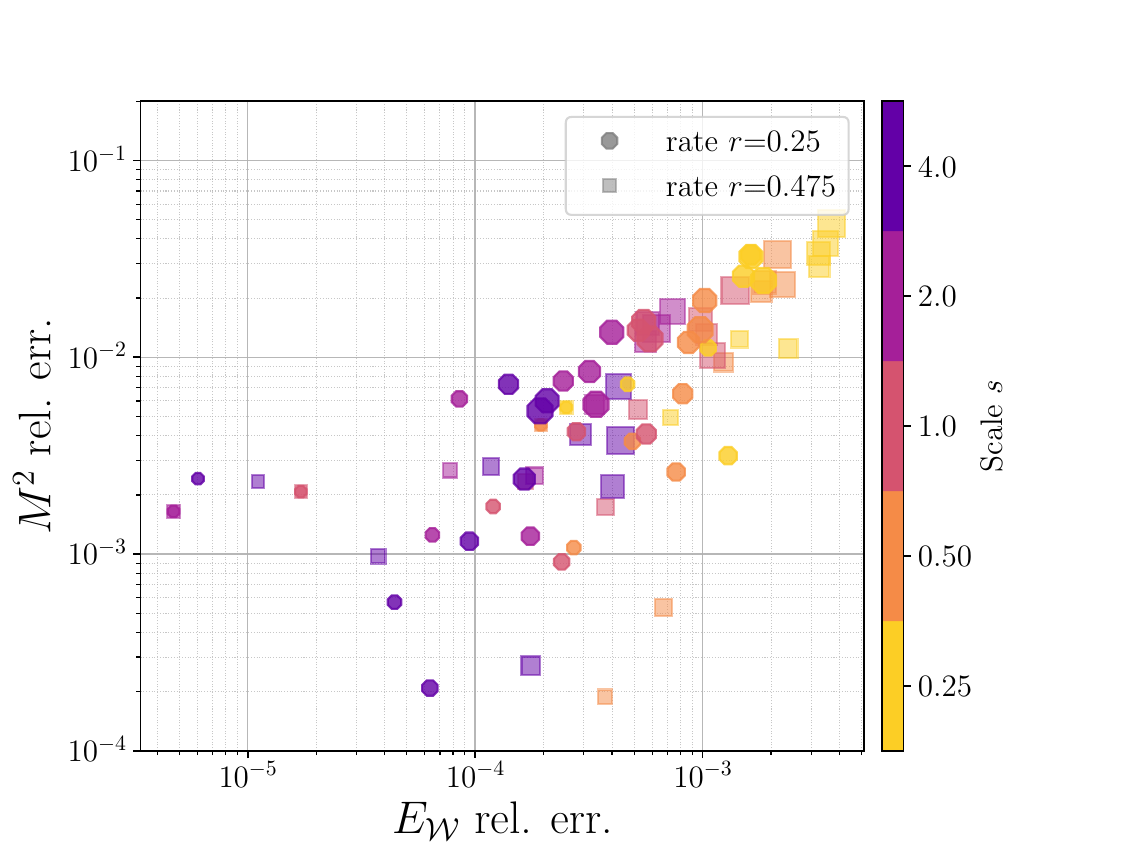}
\caption{\small The relative errors of our estimates of the sublattice magnetization determined from \Cref{eq:M} plotted against the relative errors of our variational energies. All relative errors are computed using reference values obtained from QMC~\cite{sandvik_loop_2010,anders_unpublished}. The color and marker indicate the training schedule parameters $s$ and $r$ and the size of the marker indicates the system size corresponding to each estimate.}
    \label{fig:errors}
\end{figure}

\Cref{fig:errors} shows the relative error of the finite-size squared sublattice magnetization estimates plotted against the relative error of the variational energies for the SLAHM with periodic boundary conditions for $L$ up to 20, which is the largest system size for which we computed $M^2$. We compute the relative error of the squared sublattice magnetization with respect to the QMC estimates found in Ref.~\cite{sandvik_loop_2010}. We compute the relative error of the variational energy with respect to QMC energies that will appear in a forthcoming manuscript~\cite{anders_unpublished}. 
This figure confirms that the most accurate energies also yield the most accurate estimates of the sublattice magnetization, which is consistent with the results reported in the previous sections; however, below a certain threshold, there is not a strong trend between the two relative errors. 
Other VMC studies have examined the decoupling of the accuracy of correlation functions from the accuracy of the energy~\cite{ferrari_neural_2019}. It has also been noted that achieving highly accurate correlations is more challenging than achieving high accuracy for the variational energy~\cite{becca_variational_2009}, but this does not fully explain the observed behavior.

These results show two distinct trends. First, we see that the relative error in the energy and the relative error in correlation quantities appear to scale linearly with one another for large relative errors (upper right corner in \Cref{fig:errors}). Second, we see that systematically improving the variational energy does not guarantee the same systematic improvement in the correlations that are captured. Thus, it is essential that the correlations of a state are studied just as carefully as the variational energy when aiming for highly accurate variational wavefunctions. 

\section{Conclusion}
\label{sec:conclusion}

In this work, we study the SLAHM using two-dimensional RNN wavefunctions. Taking advantage of the ability to iteratively retrain these variational ans\"{a}tze, we are able to simulate two-dimensional systems with linear lattice sizes $L$ up to 32. Using these results, we then extract accurate estimates of the ground-state energy and the sublattice magnetization in the thermodynamic limit.  We introduced a parameterized training schedule which allows us to demonstrate that our results can be systematically improved with longer training times. Even for the longest training times studied in this work ($s=4.0$ and $r=0.25$), the iterative retraining approach significantly reduces the amount of computational resources needed to simulate such large systems. 

Even though our estimates for $E_\infty$ and $M_\infty$ are already in good agreement with literature values, we would like to acknowledge a potential avenue for improvement. Many of the strongest NQS results have been achieved using Stochastic Reconfiguration (SR)~\cite{becca_quantum_2017} or its variants~\cite{chen_empowering_2024,rende_simple_2024}. SR is an optimization method that can be derived from the imaginary time evolution of a quantum state. This method updates the variational wavefunction according to gradients that are adjusted using information about the geometric structure of the optimization landscape. In this work we optimize our RNN wavefunctions using Adam: a first-order optimization method that is common in classical machine learning~\cite{kingma2017adammethodstochasticoptimization}. It is natural to wonder how our results might improve if our simulations are performed using a technique that is better-tailored to the task at hand. It has, however, been observed that SR is not effective when applied to RNNs~\cite{donatella_autoregressive,lange2024neural}. The root cause for this poor performance is not well-understood, however, it is possible that the recurrent nature of RNNs, the same property that allows us to iteratively retrain our variational wavefunctions, is to blame. The question then arises if it is possible to reconcile the two methods. 

While the practical similarity between classical machine learning tasks and NQS is an active area of research, it is worth drawing connections between our results and relevant topics in the classical machine learning literature. For instance, a significant discovery in the modern age of deep learning is the existence of ``neural scaling laws'' which seem to generally describe how training loss behaves as a function of certain scaling variables such as the number of parameters in a neural network, the size of the dataset, or the amount of compute used during training~\cite{kaplan2020scaling, hernandez2021scaling, bordelon2024feature}. Further investigation might reveal that the systematic improvement of our results could be quantified by a power law scaling of the training time used to obtain them. Recently, neural scaling laws have been observed in the context of NQS applied to quantum chemistry problems~\cite{jiang2025neural}. Ref.~\cite{jiang2025neural} demonstrates a power law decay in ground-state energy accuracy, using the number of parameters in their ans\"{a}tze as the main scaling variable. 

In the present work, we considered only one network size, $d_h=256$, which we chose based on previous iterative retraining studies~\cite{roth_iterative_2020,hibat-allah_supplementing_2024}. 
Since our RNN wavefunctions show systematic improvement with increased training time, we believe that the quality of our simulations is not limited by the expressiveness of the ans\"{a}tze. Still, the size of the hidden vectors $d_h$ is an important hyperparameter to explore further. For one, the scaling of $d_h$ could be the key to revealing the aforementioned neural scaling laws. Second, our iterative retraining experiments would provide an interesting way to study how $d_h$ should scale with the size of the system $N = L^2$. If an RNN wavefunction is not expressive enough to accurately represent the ground state for a given system size $L$, then increasing the training time should not improve the accuracy. Furthermore, increasing $L$ beyond the system size where accuracy saturates due to limitations in expressiveness should lead to an observable degradation in simulation quality. Given the connections between matrix product states and RNN wavefunctions~\cite{wu_tensor,yang_when_2024}, it is possible that there exists a physically meaningful way to scale $d_h$ with system size $N$. We leave such experiments for future work.

It is also possible to look to machine learning literature for insights into the success of iterative retraining.
It is understood that the recurrent nature of RNNs, which allows for weight sharing, is the property that permits iterative retraining; however, the more subtle question is why this technique is successful. Indeed, it is not always the case that neural networks trained to represent ground states are able to generalize in system size~\cite{fitzek_rydberggpt_2024}. 
What is learned about the ground state of one finite-size system that is transferrable to the ground state of the next system size?
The fact that our RNN wavefunctions generalize from one system size to the next suggests that they are ``feature learning'', or learning non-trivial patterns in the data rather than memorizing the data. Features that would be transferrable between different system sizes could be e.g. correlations. The generalization properties of NQS are the focus of ongoing research~\cite{westerhout_many-body_2023,moss2025double}.

Ultimately, this work serves to carefully benchmark the performance of iteratively retrained RNN wavefunctions for a well-understood problem. In \Cref{app:compare}, we directly compare all of our estimated quantities with values from the literature, showing the success of our methods. Our results demonstrate that RNNs are a powerful tool capable of accurately representing ground states for finite-size systems and providing reliable estimates of ground-state observables in the thermodynamic limit. 
Having proved the effectiveness of these methods, the next step is to perform similar simulations for the more challenging triangular lattice spin-$\frac{1}{2}$ antiferromagnetic Heisenberg model (TLAHM)~\cite{moss2025leveraging_tri}.

\section{Code Availability}
\label{sec:code}
Our implementation of the presented methods, all scripts needed to reproduce our results, and the data needed to reproduce the figures in this manuscript are openly available on GitHub \url{https://github.com/mschuylermoss/HeisenbergRNN}~\cite{Moss2025git}. Our code relies on TensorFlow~\cite{tensorflow2015-whitepaper}, NumPy~\cite{harris2020array}, and Matplotlib~\cite{Hunter:2007}.

\section{Acknowledgments}
\label{sec:acknowledgments}
SM would like to acknowledge Chris Roth for helpful discussions throughout the completion of this work. SM would also like to thank Anders Sandvik for sharing improved QMC estimates of observables for finite-size ground states. The improved estimates will appear in a forthcoming manuscript~\cite{anders_unpublished}.

We acknowledge financial support from the
Natural Sciences and Engineering Research Council of
Canada (NSERC) and the
Perimeter Institute. Research at Perimeter Institute is
supported in part by the Government of Canada through
the Department of Innovation, Science and Economic
Development Canada and by the Province of Ontario
through the Ministry of Economic Development, Job
Creation and Trade. RW acknowledges support from the Flatiron Institute. The Flatiron Institute is a division of the Simons Foundation. 

This work involved many large-scale simulations.
SM and RW thank the Flatiron Institute Scientific Computing Center for computational resources and technical support. Resources used in preparing this research were provided, in part, by the Province of Ontario, the Government of Canada through CIFAR, and companies sponsoring the Vector Institute \url{https://vectorinstitute.ai/#partners}. Additional resources were provided by the Shared Hierarchical Academic Research Computing Network (SHARCNET) and
the Digital Research Alliance of Canada.

\appendix

\section{Marshall-Peierls Sign Rule}
\label{app:marshallsign}

The square lattice is bipartite, meaning there are two independent sublattices $A,B$ with $N/2$ spins on each sublattice if $L$ is even. 
One can define the following local unitary, 
\begin{align*}
    \mathcal{U} = \text{exp}\left( 
    -\text{i}\pi\sum_{b\in B}\hat{S}_b^{z}
    \right),
\end{align*}
which rotates all spins on sublattice B by $\pi$ around the z-axis.
Applying this unitary to our Hamiltonian defined in \Cref{hamiltonian}, we find that the off-diagonal components take on a leading factor of $-1$, 
\begin{align}
    \hat{H}^\prime &= \mathcal{U}^\dagger\hat{H}\mathcal{U} \nonumber\\
    &= \sum_{\langle ij \rangle} \left[\hat{S}^z_i\hat{S}^z_j - (\hat{S}^x_i\hat{S}^x_j + \hat{S}^y_i\hat{S}^y_j)
    \right].
    \label{mpsr_hamiltonian}
\end{align}
When the off-diagonal components of a Hamiltonian are negative, then the Hamiltonian is called stoquastic and the Perron-Frobenius theorem implies that the ground-state wavefunction will have real and positive amplitudes~\cite{perron_zur_1907,frobenius_uber_1912}.
This local unitary does not change the value of the ground-state energy or other local observables.  
Therefore, instead of finding the ground state of the original Hamiltonian, one can instead find the ground state of the stoquastic Hamiltonian defined in \Cref{mpsr_hamiltonian}. This is a simpler task because one does not need to learn a non-trivial sign structure.

\section{Details of RNN Wavefunctions}
\label{app:rnn_details}

As mentioned, the cell of an RNN is the main information processing unit. We use gated recurrent units (GRUs), which are known to alleviate the vanishing and exploding gradients problem that RNNs face~\cite{cho2014learning,chung2014empirical}. The GRUs we employ include the computation of three quantities: a candidate hidden vector $\Tilde{\vec{h}}_{i,j}$, an update gate $\vec{u}_{i,j}$, and finally the output hidden vector $\vec{h}_{i,j}$. The calculation of $\Tilde{\vec{h}}_{i,j}$ and $\vec{u}_{i,j}$ are different depending on the boundary conditions being considered. A summary of the parameters for each type of RNN wavefunction is listed in \Cref{tab:parameters}.

For systems with periodic boundary conditions, we include pseudo-periodic boundary conditions into our two-dimensional RNN wavefunction~\cite{luo_gauge-invariant_2023, hibat-allah_investigating_2023}. In this case, the cell takes four hidden vectors as inputs. We employ a regular GRU cell. The input to this cell are the four hidden vectors $\vec{h}$ and one-hot representations of a single spin in the full configuration $\vec{\sigma}_i$ obtained from the nearest-neighbor RNN cells. Both hidden vectors and input spins are initialized with zeros and get assigned values as we progress through the autoregressive sequence. Therefore, only hidden vectors and spins coming from nearest-neighbors that come earlier in the autoregressive sequence will contribute to the calculations in the cell, thus satisfying the requirements of  \Cref{chain_rule_of_prob}.
The RNN cell computes a  candidate hidden vector and update gate as follows:
\begin{align*}
    \Tilde{\vec{h}}_{i,j} &= \text{tanh}\left(
    \left[\vec{h}_{\text{input}}; \vec{\sigma}_{\text{input}}]\right]
    \mathbf{W} + \vec{b}\right),
    \\
     \vec{u}_{i,j} &= \text{sigmoid}\left(
     \left[\vec{h}_{\text{input}}; \vec{\sigma}_{\text{input}}]\right]
    \mathbf{W}_g + \vec{b}_g\right).
\end{align*}
with
\begin{align*}
    \vec{h}_{\text{input}} &= [\vec{h}_{i-1,j};\vec{h}_{i,j-1};\vec{h}_{i+1,j};\vec{h}_{i,j+1}],\\
    \vec{\sigma}_{\text{input}} &= [\vec{\sigma}_{i-1,j};\vec{\sigma}_{i,j-1};\vec{\sigma}_{i+1,j};\vec{\sigma}_{i,j+1}],
\end{align*}
where $[\vec{a},\vec{b}]$ denotes the concatenation of two vectors. Note that for periodic boundary conditions, we have $\vec{h}_{i, L+1}\equiv \vec{h}_{i, 1}$ and $\vec{h}_{L+1, j}\equiv \vec{h}_{1, j}$ as well as $\vec{\sigma}_{i, L+1}\equiv \vec{\sigma}_{i, 1}$ and $\vec{\sigma}_{L+1, j}\equiv \vec{\sigma}_{1, j}$. 

For our simulations of systems with open boundary conditions, which can be found in \Cref{app:OBC}, we use a Tensorized version of the GRU cell, which improves the expressiveness of the ansatz~\cite{hibat-allah_variational_2021} at the cost of a larger number of parameters. This cell only accepts the two hidden vectors and spin states that come from the nearest-neighbor cells that appear previously in the autoregressive sequence. The candidate hidden vector and update gate are computed as follows:
\begin{align*}
    \Tilde{\vec{h}}_{i,j} &= \text{tanh}([\vec{\sigma}_{i-1,j};\vec{\sigma}_{i,j-1}]
    \mathbf{T} [\vec{h}_{i-1,j};\vec{h}_{i,j-1}] + \vec{b}),
    \\
    \vec{u}_{i,j} &= \text{sigmoid}([\vec{\sigma}_{i-1,j};\vec{\sigma}_{i,j-1}]
    \mathbf{T}_g [\vec{h}_{i-1,j};\vec{h}_{i,j-1}] + \vec{b}_g).
\end{align*}

For both the linear and tensorized GRU cell, we compute an output hidden vector from the candidate hidden vector $\Tilde{\vec{h}}_{i,j}$ and the update gate $\vec{u}_{i,j}$,
\begin{align*}
    \vec{h}_{i,j} = \vec{u}_{i,j} \odot\Tilde{\vec{h}}_{i,j} + (1 - \vec{u}_{i,j}) \odot \left([\vec{h}_{i-1,j};\vec{h}_{i,j-1}]
    \mathbf{W}_{\text{merge}}\right).
\end{align*} 
Thus, the update gate serves to modulate how much the output hidden vector changes from the input hidden vectors.

\begin{table}
    \centering
    \begin{tabular}{c|c|c}
    \hline\hline
         & PBC & OBC \\
         \hline
        cell & $\mathbf{W} \in \mathbb{R}^{4d_h+4d_\sigma \times d_h}$ & $\mathbf{T} \in \mathbb{R}^{2d_h \times 2d_\sigma \times d_h}$ \\
     & $\vec{b} \in \mathbb{R}^{d_h}$ & $\vec{b} \in \mathbb{R}^{d_h}$ \\
     & $\mathbf{W}_g \in \mathbb{R}^{4d_h+4d_\sigma \times d_h}$ & $\mathbf{T}_g \in \mathbb{R}^{2d_h \times 2d_\sigma \times d_h}$ \\
     & $\vec{b}_g \in \mathbb{R}^{d_h}$ & $\vec{b}_g \in \mathbb{R}^{d_h}$ \\
     & $\mathbf{W}_{\text{merge}} \in \mathbb{R}^{4d_h  \times d_h}$ & $\mathbf{W}_{\text{merge}} \in \mathbb{R}^{2d_h \times d_h}$ \\\hline
dense layer & $\mathbf{U} \in \mathbb{R}^{d_h \times d_\sigma}$ & $\mathbf{U} \in \mathbb{R}^{d_h \times d_\sigma}$ \\
     & $\vec{c} \in \mathbb{R}^{d_\sigma}$ & $\vec{c} \in \mathbb{R}^{d_\sigma}$ \\\hline
total $\#$ & 791,554 & 1,180,674 \\
        \hline\hline
    \end{tabular}
    \caption{Details of the parameters for both types of RNN wavefunctions. $d_h$ is the size of the hidden vector $\vec{h}$ and $d_\sigma$ is the dimension of the Hilbert space for a single spin. For a spin-$\frac{1}{2}$ system, $d_\sigma=2$. For all RNN wavefunctions used in this work, we fix $d_h=256$.}
    \label{tab:parameters}
\end{table}

This output vector $\vec{h}_{i,j}$ is then used to compute the conditional probability over $\sigma_{i,j}$ using a final dense layer with a Softmax activation function:
\begin{align*}
    p_\mathcal{W} (\sigma_{i,j}\vert\sigma_{<i,j}) = \text{Softmax}(\mathbf{U}\vec{h}_{i,j} + \vec{c}) \cdot \vec{\sigma}_{i,j}.
\end{align*}
The overall wavefunction amplitude corresponding to the full spin configuration ${\vec{\sigma}}$ is $\Psi({\vec{\sigma}}) = \prod_{i,j} \sqrt{p_\mathcal{W}(\sigma_{i,j})}$.

For spin-$\frac{1}{2}$ systems $d_\sigma=2$. For all of the simulations presented in this work we fix $d_h=256$, which seems sufficiently expressive for the largest systems reached with our iterative retraining. 

\section{Training Details}
\label{app:training_details}

We start our training procedure by optimizing an RNN wavefunction for the SLAHM with $L=6$, where we can achieve highly accurate results. This training is performed in three stages. 
In the first stage, we use a fixed learning rate of $\gamma = 5\times10^{-4}$. 
In the second and third stage, we decay the learning rate using 
\begin{align*}
    \gamma (t) = \gamma_0 \times (1 + (t/\delta))^{-1},
\end{align*}
with $\gamma_0 = 5 \times 10^{-4}$ and  $\delta = 5\times10^3\times s$. Note that $\gamma (t)$ is always $\gamma \approx 5\times10^{-5}$ at the end of the third stage of training, independent of the choice of $s$ in \Cref{eq:schedule}.

To control the number of training steps we use \Cref{eq:schedule}, which depends on two constants $C$ and $F$, the rate $r$, and the scale $s$ (see also \Cref{fig:steps_per_L} in the main text). Due to the form of \Cref{eq:schedule}, the bulk of the training takes place in stages 1 and 2. Since symmetry averaging significantly increases the computational cost, we only apply the $C_{4v}$ symmetry from stage 3 onward. For very large system sizes $N_{\text{steps}}\approx s\times F$ and we fix $F=2\times10^3$.

After we have converged to a solution for $L=6$, we iteratively retrain our RNN wavefunction for $L>6$. We fix $C=101\times 10^3$, $\gamma = 5\times10^{-5}$, and we always enforce lattice symmetries. At all stages of training, we enforce the $U(1)$ symmetry.
We summarize the training schedule in \Cref{tab:training}.

\begin{table}[h]
    \centering
    \begin{tabular}{c|c|c|c|c}
    \hline\hline
        System size & $C$ & $\gamma$ & $C_{4v}$ & $U(1)$ \\\hline
        $L=6$ (stage 1) & $10^3$ + $50\times10^3$ & $5\times10^{-4}$ & False & True\\
        $L=6$ (stage 2)  & $76 \times10^3$ & $\gamma(t)$ & False & True\\
        $L=6$ (stage 3)  & $101\times10^3 $& $\gamma(t)$ & True & True\\
        $L>6$ & $101\times10^3 $ & $5\times10^{-5}$ & True & True\\
    \hline\hline
    \end{tabular}
    \caption{\small Hyper parameters for the iterative retraining of RNN wavefunctions.}
    \label{tab:training}
\end{table}

\section{Runtime Details}
\label{app:runtimes}

\begin{figure}
    \centering
    \includegraphics[width=\linewidth]{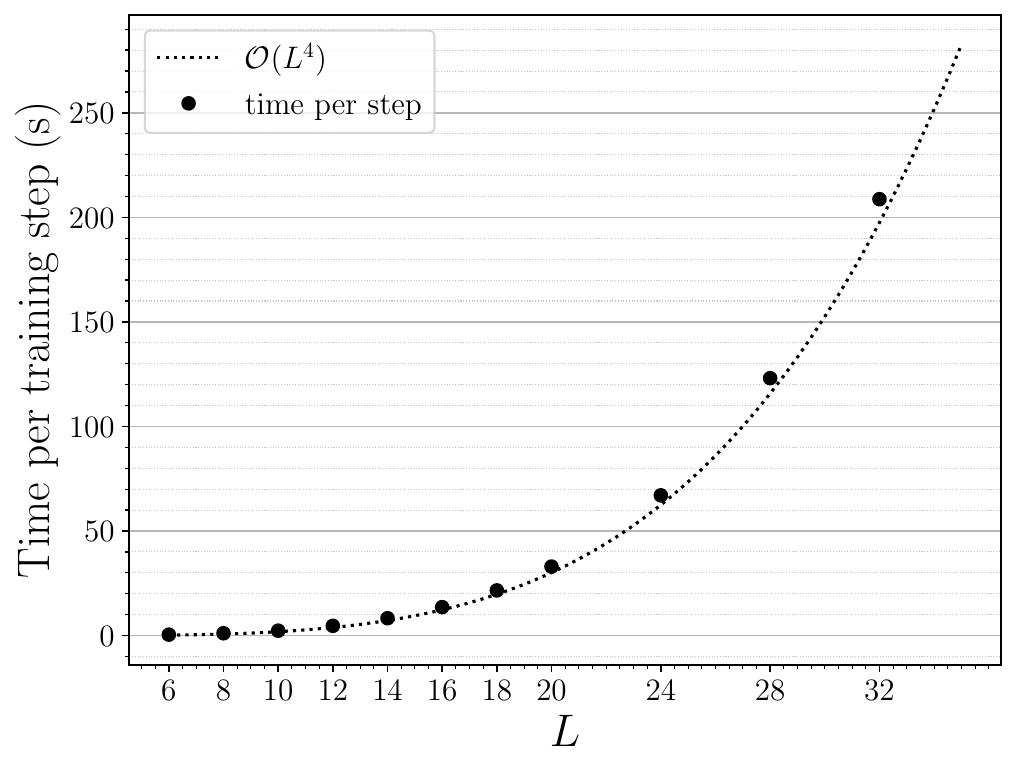}
    \caption{\small The amount of time in seconds it takes to complete one train step for a given system size with periodic boundary conditions using four A100 GPUs.}
    \label{fig:time_per_step}
\end{figure}

There is a one-to-one correspondence between the cells in the RNN wavefunction and the sites on the lattice. As a result, one forward pass of the RNN wavefunction scales with the number of the spins in the system $\mathcal{O}(N)$, since both sampling and inference must be performed sequentially along the autoregressive path. The dominant part of a single training step is the computation of the variational energy, where one has to obtain the log amplitudes of all configurations that are connected to the input samples by the Hamiltonian. The number of connected configurations scales as $\mathcal{O}(N)$ for the square lattice with nearest neighbor interactions. Therefore, for a given training step, one must make $\mathcal{O}(N)$ forward passes of the RNN which costs $\mathcal{O}(N)$ for each pass. Thus, the overall time per step should scale as $\mathcal{O}(L^4)$, since $N=L^2$. \Cref{fig:time_per_step} shows how the time per step scales in practice and how it compares to this theoretical estimate. We emphasize that weight sharing keeps the size of the RNN wavefunction constant throughout the iterative retraining, so that the increase in runtime is solely due to the increasing system size $L$.
\begin{figure}
    \centering
    \includegraphics[width=\linewidth]{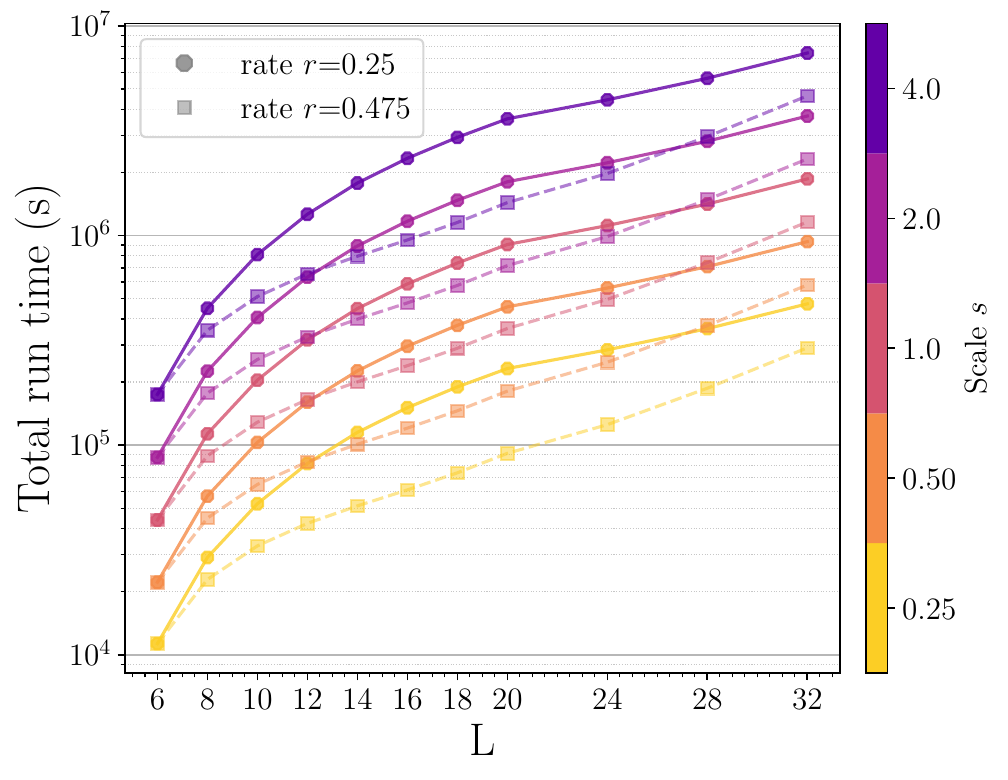}
    \caption{\small The amount of time in seconds it takes to complete the training up to a given system size with periodic boundary conditions using four A100 GPUs.}
    \label{fig:run_times}
\end{figure}

Our goal in decaying the number of training steps with increasing system size is to counteract the rapidly increasing time per step shown in \Cref{fig:time_per_step}. \Cref{fig:run_times} shows the total run times for the 10 training schedules we consider. The run time shown is the time it takes to perform training for all lattice sizes up to and including a given $L$. 

In practice, one expects the optimization of a neural network quantum state to become more challenging for larger physical system sizes if one is optimizing from scratch, due to the larger Hilbert space. If this is the case, then for simulations of large systems one would likely need to train longer (for more training steps) in order to converge to a good approximation of the ground state. Given that the time per training step grows rapidly with system size, as seen in \Cref{fig:time_per_step}, increasing the number of steps with $L$ would lead to a rapid increase in total run time. Iterative retraining and the ability to \emph{decrease} the number of training steps as the system size $L$, and consequently the time per training step, increases, allow us to dramatically reduce the amount of computation time needed to obtain simulation results for many large system sizes. 

Importantly, we do not study how the choice of $d_h$ should scale with $N$, so we are unable to comment on how the run time would scale if RNN wavefunctions of different size were used for physical systems of different size.

\section{V-Score Scaling}
\label{app:vscore}

Recently, Wu et. al.~\cite{wu_variational_2024} introduced a new metric for quantifying the accuracy of a variational method. The idea was to define a quantity which is intensive (e.g. independent of system size) and takes into account the variational energy and the variance of the variational energy, both of which are used to assess the quality of variational result. The introduced metric, which is called the V-score, can be defined as 
\begin{align*}
    \text{V-score} := \frac{N\,\text{Var}(E)}{(E- E_{\infty})^2},
\end{align*}
where $E_\infty$ is the zero-point energy, which is zero for the Heisenberg model.
\begin{figure}
    \centering
    \includegraphics[width=\linewidth]{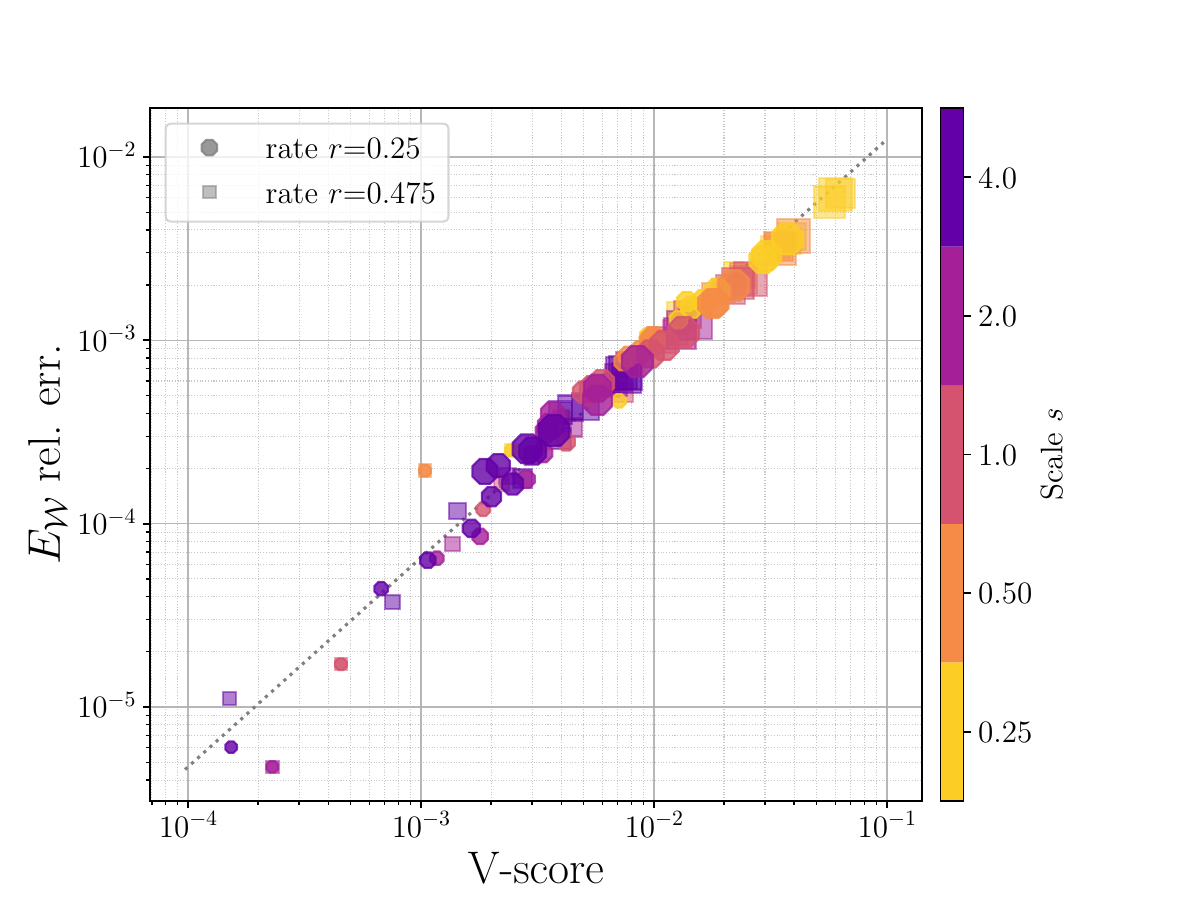}
    \caption{\small The V-score and relative energy error obtained from our final variational energies plotted against one another. Relative energy errors are computed with respect to energies from QMC. The color and marker indicate the training schedule parameters $s$ and $r$ used to obtain that result. The size of the marker indicates the system size for each quantity shown. The dotted line shows a linear fit through the log-scaled data as a visual guide to the linear nature of the relationship between the two quantities.}
    \label{fig:vscores_errors}
\end{figure}

To further motivate this quantity, the authors in~\cite{wu_variational_2024} show that the V-scores for various models scale linearly with the corresponding relative errors of the energies. For the SLAHM with periodic boundary conditions, we have reference energies from QMC that are numerically exact~\cite{anders_unpublished}, so we can examine this behavior for our own results. \Cref{fig:vscores_errors} shows the relative error of our variational energies plotted against the corresponding V-scores. Indeed we find a linear relationship between these two quantities, even for large $L$ where some energy relative errors begin to diverge as seen in \Cref{fig:energies_peri}(a).

\begin{figure}
    \centering
    \includegraphics[width=\linewidth]{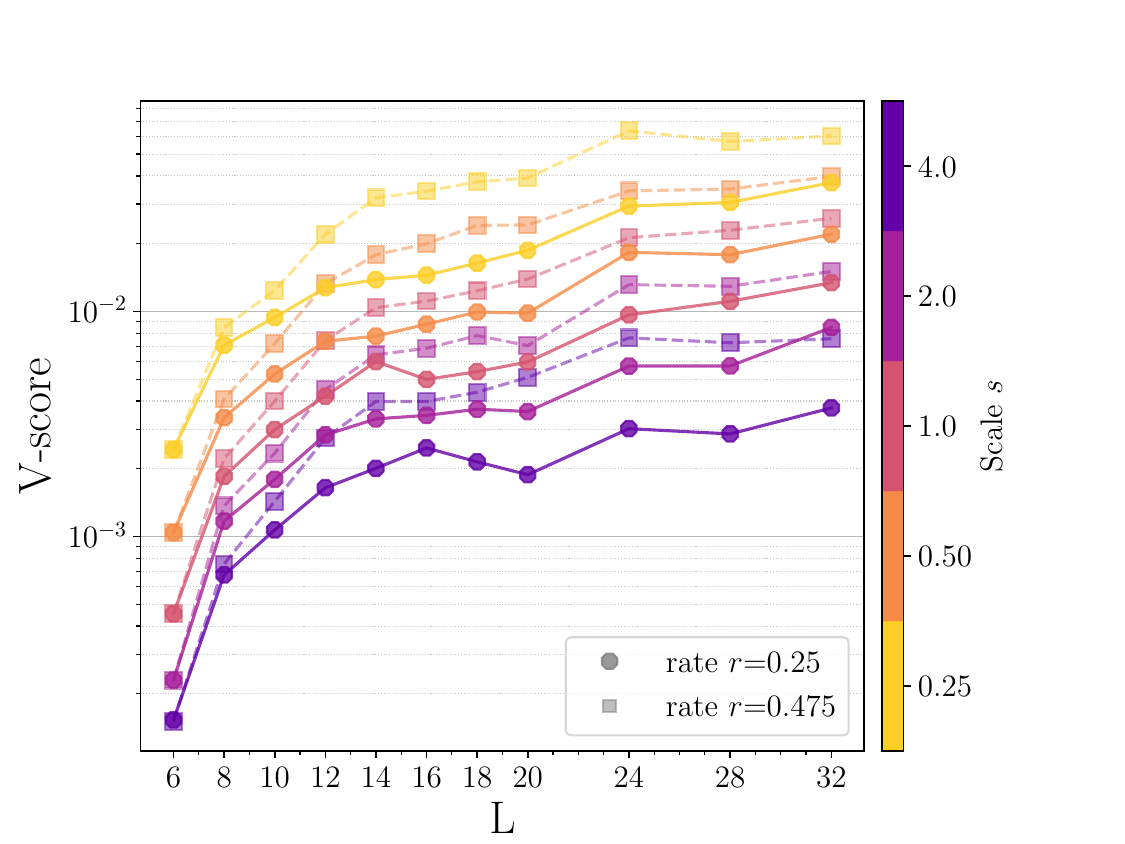}
    \caption{\small V-scores obtained for all lattice sizes $L$ from all of our RNN wavefunctions. Each RNN wavefunction was trained with different training schedules parameterized by $s$ and $r$. }
    \label{fig:vscores}
\end{figure}
As mentioned, the SLAHM on lattices with $L=6$ and $L=10$ are two common benchmarks for variational methods. As such, there are many recorded V-scores. Figures 3 and 4 in ref.~\cite{wu_variational_2024} show the range of V-scores obtained for the SLAHM on these lattice sizes with both open and periodic boundary conditions. Our V-score results for these system sizes are in agreement with the values seen there. Interestingly, the reported V-scores that are superior to the V-scores we obtain are also from simulations with a different type of RNN wavefunction (multi-layer LSTM). Because we perform simulations for system sizes much larger than those cataloged in ref.~\cite{wu_variational_2024}, we report the V-scores for our results for the SLAHM with periodic boundary conditions in \Cref{fig:vscores}. These results further confirm that our results can be systematically improved through longer training times as determined by a larger scale $s$ and a slower rate $r$.

\section{Zero-variance extrapolation of the variational energy}
\label{app:zero-var}

\begin{figure}
    \centering
    \includegraphics[width=\linewidth]{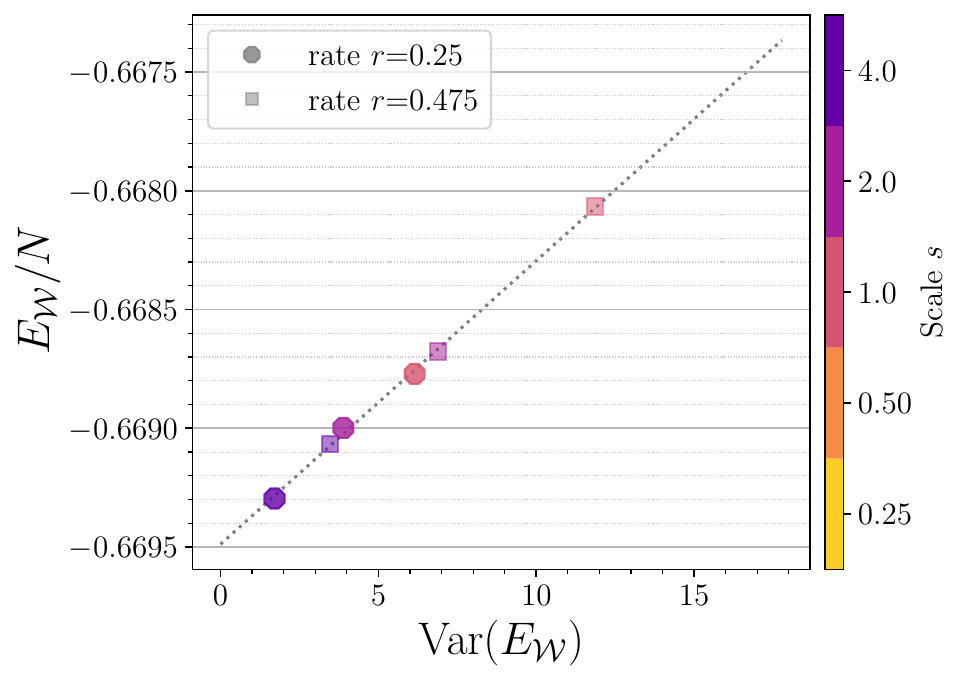}
    \caption{The zero-variance extrapolation of our best variational energies from \Cref{fig:energies_peri} for $L=32$. Each variational energy comes from an RNN wavefunction trained according to the schedule given by \Cref{eq:schedule} and a unique choice of the scale $s$ and rate $r$. The dotted grey line shows the linear fit through the data. }
    \label{fig:zero-var-extrap}
\end{figure}

Using a sequence of systematically improving variational states, it is common to extrapolate the corresponding variational energies to their zero-variance limit~\cite{nomura_restricted_2017,kwon_effects_1998,becca2000stabilitydwavesuperconductivitytj,Sorella2001,hu_direct_2013,fu_variance_2023}. This type of extrapolation is motivated by the zero-variance principle~\cite{becca_quantum_2017}. \Cref{fig:zero-var-extrap} shows an example of this type of extrapolation for our best variational energies for $L=32$ and periodic boundary conditions. As mentioned, we perform this fit only through our best variational energies, which come from simulations with $s\geq1.0$.

\section{Results for the SLAHM with open boundary conditions}
\label{app:OBC}

Here we present our results for the SLAHM with open boundary conditions. 
The first quantities to assess are the variational energies from our simulations and their corresponding variances. 

\begin{figure}
    \includegraphics[width=\linewidth]{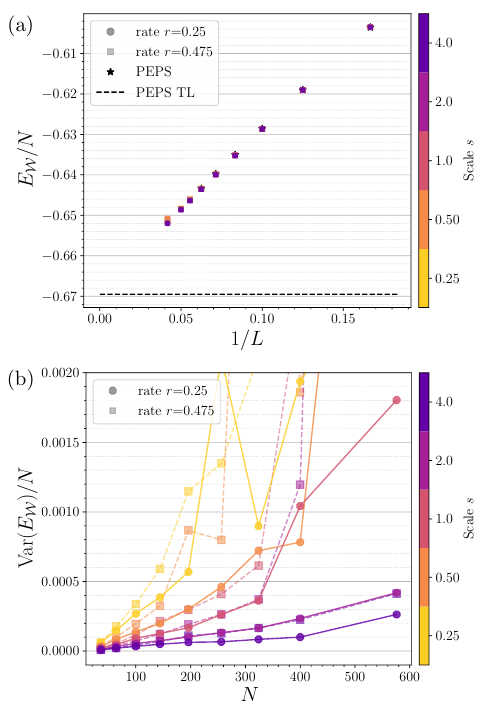}
    \caption{\small 
    (a) The final variational energy obtained from all optimized RNN wavefunctions for each system size. All of the final variational energies were estimated with 10k samples. Reference energies and the thermodynamic limit (TL) value of the ground-state energy per spin obtained from PEPS simulations are shown for reference.
    (b) The variance of the final variational energy obtained from all optimized RNN wavefunctions for each system size. 
    }
    \label{fig:energy_variance_open}
\end{figure}

\Cref{fig:energy_variance_open}(a) shows that our variational energies are in very close agreement with reference values obtained from gradient-optimized PEPS simulations~\cite{liu_gradient_2017}. In some cases, our variational energies are lower than the reported PEPS values, which is consistent with previous works~\cite{hibat-allah_supplementing_2024}. Unlike what we observed with the simulations of the SLAHM with periodic boundary conditions, none of the variational energies strongly diverge from the reference values. 
\Cref{fig:energy_variance_open}(b) displays the variances of our variational energies. 
Although the variational energies seem to be in good agreement across all simulations, the variances still improve systematically based on the training schedule. Training with a larger scale $s$ or a slower rate $r$, and thus lengthening the total training time, improves the variance of the variational energy. 
Interestingly, these variances also hint at some instability in the training for larger system sizes when the scale $s$ is smaller. This suggests that, at least in the case of open boundary conditions, the RNN wavefunctions for a given system size $L$ must be sufficiently trained before moving on the next system size. Only the three simulations with the longest training times maintain low variances across all system sizes.
In general, these variances are much lower than the variances obtained from our simulations of the SLAHM with periodic boundary conditions. 

\begin{figure}
    \includegraphics[width=\linewidth]{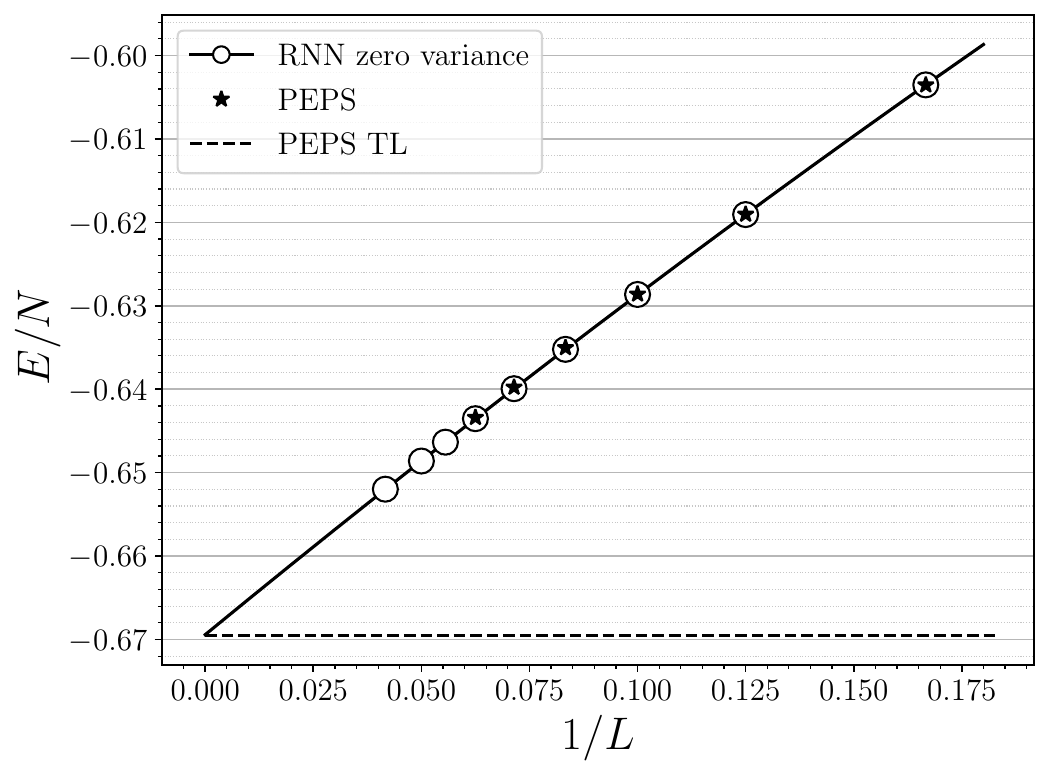}
    \caption{\small
    Energy estimates obtained from extrapolating the variational energies to their zero-variance value compared to reference energies obtained from PEPS simulations.  The thermodynamic limit (TL) value of the ground-state energy per spin obtained from PEPS is shown for reference.}
    \label{fig:zero_var_energy_open}
\end{figure}

We perform zero-variance extrapolation using only the results from the simulations that consistently produce variational energies with low variance. This is because the zero-variance extrapolation relies on having systematically improving results. The zero-variance energies are then used to perform a finite-size scaling of the energy. 
For the SLAHM with open boundary conditions, we fit a second-order polynomial in $1/L$ to our energies to extract the ground-state energy in the thermodynamic limit~\cite{liu_gradient_2017}.
The scaling behavior is different from that of the energies of the SLAHM with periodic boundary conditions because finite-size effects are larger for systems with open boundary conditions. 
\Cref{fig:zero_var_energy_open} shows our zero-variance energies for the SLAHM with open boundary conditions and the finite-size scaling.
Using this second-order fit and a wild bootstrap~\cite{wu_bootstrap} method with $10^3$ bootstraps, we obtain an extrapolated ground-state energy of $E_{\infty}=-0.669425(1)$, which is in close agreement with the reported PEPS value $E_\infty^{\text{PEPS}} = -0.66948(42)$~\cite{liu_gradient_2017} and also with the value obtained from QMC $E_\infty^{\text{QMC}}=-0.669437(5)$. While the QMC value was obtained from simulations of periodic systems, the extrapolated energies should coincide in the thermodynamic limit. 

\begin{figure}
    \includegraphics[width=\linewidth]{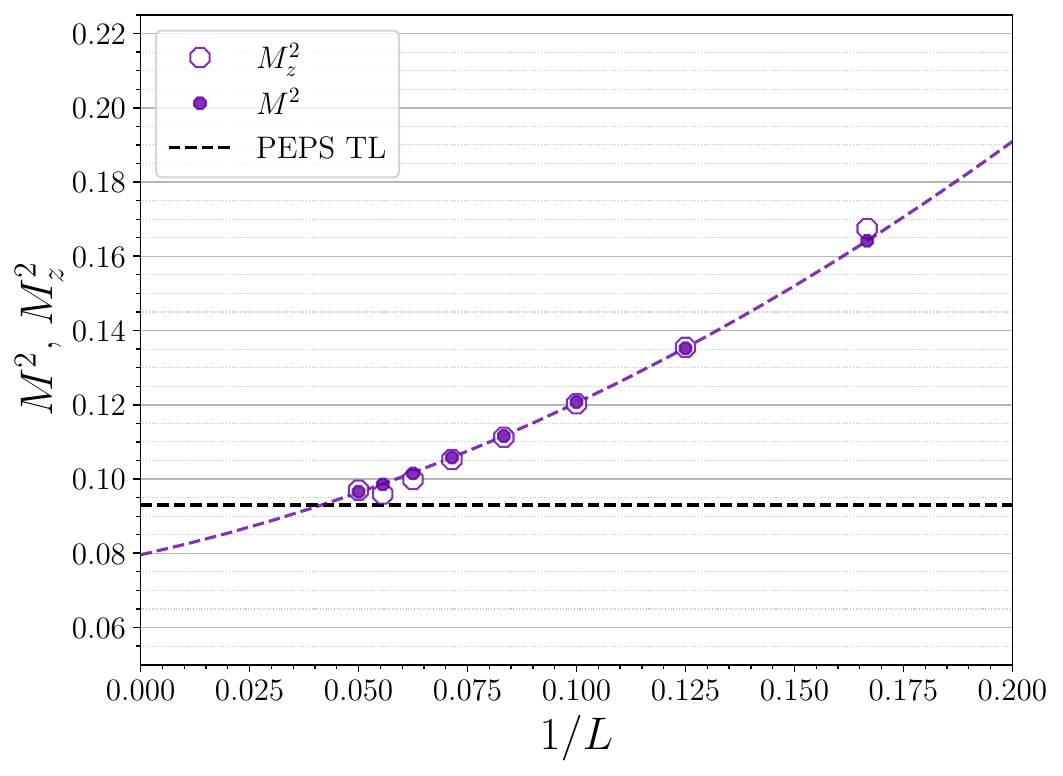}
    \caption{\small
    Estimates of the squared sublattice magnetization scaled as a function of $1/L$. These estimates of $M^2$ and $M_z^2$ were obtained from our simulations with the longest amount of training time, i.e. $
    s=4.0$ and $r=0.25$. Closed markers represent the values of $M^2$ and $M_C^2$ estimated using the correlations defined by \Cref{eq:real_space_corr}, while open markers represent the values of $M^2$ and $M^2_C$ that are estimated using the correlations defined by \Cref{eq:real_space_corr_z}. The thermodynamic limit (TL) value of the squared sublattice magnetization obtained from PEPS is shown for reference.
    }
    \label{fig:magnetization_open}
\end{figure}

We also examine estimates of the sublattice magnetization obtained from our simulations of  open systems with $L$ up to 20. In particular, we show estimates of the sublattice magnetization obtained from the structure factor, as defined in \Cref{eq:M}. We show both estimates of this value: one obtained using the full correlations defined in \Cref{eq:real_space_corr} and one obtained using only the $z$ basis correlations defined in \Cref{eq:real_space_corr_z}. 
\Cref{fig:magnetization_open} shows our estimates of these values obtained from our longest and most accurate simulation, i.e. with the largest scale $s=4.0$ and the slowest rate $r=0.25$. We also perform a finite-size scaling of these values using a second-order polynomial in $1/L$. Liu \emph{et al.}~\cite{liu_gradient_2017} performed a similar finite-size scaling, but only reference values for their extrapolated value of $M^2$ were provided. The extrapolated value obtained from our finite-size scaling is reduced from the value obtained from the PEPS simulation, and reduced from the value obtained from QMC~\cite{sandvik_loop_2010}. This is likely due to finite-size effects which are stronger for systems with open boundary conditions. Indeed, the extrapolated value of $M_\infty^{\text{PEPS}}$ was obtained from values of $M^2$ that were estimated by considering only the smaller, central portion of the lattice precisely to reduce edge effects. 

One key detail in \Cref{fig:magnetization_open} is the lack of SU(2) symmetry breaking which was very pronounced in our sublattice magnetization results for periodic systems. It is likely that the preservation of the full SU(2) symmetry is related to the improved accuracy of the variational results as seen in \Cref{fig:energy_variance_open}. However, it is unclear why our RNN wavefunctions are able to reach these better variational results for the SLAHM with open boundary conditions.  

\clearpage
\onecolumngrid

\section{Comparison of results to benchmarks and other methods}
\label{app:compare}

In this appendix we directly compare our results to the existing literature. First, in \Cref{tab:qmc_energies} we compare our finite-size variational energies for the SLAHM with periodic boundary conditions against the reference energies from QMC~\cite{anders_unpublished}. In \Cref{tab:qmc_sublattice_magnetizations} we compare our finite-size estimates of the squared sublattice magnetization against QMC reference values~\cite{sandvik_loop_2010}. Next, in \Cref{tab:nqs_energies} we present the best variational energies obtained with other NQS architectures for $L=6$ and $L=10$ with both open and periodic boundary conditions. Finally, in \Cref{tab:TL} we compare our estimates of the ground-state energy per spin and the sublattice magnetization in thermodynamic limit (i.e. obtained from our finite-size scaling) against the existing literature. Notably, we provide the first thermodynamic limit estimates obtained with NQS.

\begin{table*}[h!]
    \centering
    \begin{tabularx}{\textwidth}{C L L}
    \hline\hline
        system size & RNN & QMC~\cite{anders_unpublished}\\
        $L$ & zero-variance $E/N$ & $E/N$ \\\hline
        6 &  $-0.67887177(7)$ & $-0.678872134(45)$ \\ 
        8 &  $-0.6735047(2)$ & $-0.67349002(4)$ \\ 
        10 & $-0.6715950(5)$ & $-0.67155264(3)$ \\ 
        12 & $-0.6707008(1)$ & $-0.67068193(3)$ \\ 
        14 & $-0.6703092(5)$ & $-0.670232291(35)$ \\ 
        16 & $-0.6699687(9)$ & $-0.66997664(3)$ \\ 
        18 & $-0.6697819(5)$ & $-0.66982039(4)$ \\ 
        20 & $-0.6697420(6)$ & $-0.66971969(4)$ \\ 
        24 & $-0.6696127(5)$ & $-0.66960443(4)$ \\ 
        28 & $-0.6695580(4)$ & $-0.66954498(4)$ \\ 
        32 & $-0.6694877(4)$ & $-0.66951136(4)$\\ 
        \hline\hline
    \end{tabularx}
    \caption{Variational energies obtained from our RNN wavefunctions after performing zero-variance extrapolation for the SLAHM with periodic boundary conditions. Improved statistical estimates of these energies and their corresponding errors were obtained using the wild bootstrap method~\cite{wu_bootstrap} and $10^3$ bootstraps. We acknowledge that many of these zero-variance energies are below the reference values, but note that this is an artifact of the zero-variance extrapolation and not a violation of the variational principle. None of the variational energies obtained from our simulations were less than the reference values. }
    \label{tab:qmc_energies}
\end{table*}

\begin{table*}[h!]
    \centering
    \begin{tabularx}{\textwidth}{C L L L L}
    \hline\hline
     system size & RNN & QMC ~\cite{sandvik_finite-size_1997}& RNN & QMC~\cite{sandvik_finite-size_1997}  \\
     $L$ & $S(L,\vec{q} = (\pi,\pi))/L^2$ & $S^{(z)}(L,\vec{q} = (\pi,\pi))/L^2$ & $C(L/2,L/2)$ & $C^{(z)}(L/2,L/2)$\\ \hline
        6  & 0.210(4) & -- & 0.1523(9) & -- \\ 
        8  & 0.178(4) & 0.177843(1) & 0.1370(7)  & 0.137595(2) \\ 
        10 & 0.159(4) & 0.159372(2) & 0.1283(6) & 0.128552(2) \\ 
        12 & 0.147(4) & 0.147448(2) & 0.1230(4)  & 0.122586(2) \\ 
        14 & 0.140(4) & 0.139153(2) & 0.1193(4)  & 0.118380(2) \\ 
        16 & 0.133(4) & 0.133067(2) & 0.1154(4)  & 0.115263(2) \\ 
        18 & 0.129(4) & 0.128412(2) & 0.1139(3)  & 0.112857(2) \\ 
        20 & 0.125(4) & 0.124748(2) &  0.1117(2) & 0.110954(2) \\ 
        24 & -- & 0.119350(2) & 0.1089(2) & 0.108125(2) \\ 
        28 & -- & 0.115573(2) & 0.1085(2) & 0.106126(2) \\ 
        32 & -- & 0.112782(2) & 0.1060(2) & 0.104636(2) \\ 
        \hline\hline
    \end{tabularx}
    \caption{Sublattice magnetization estimates from our best simulations for the SLAHM with periodic boundary conditions. These estimates were obtained from the RNN wavefunction trained using a training scheduled defined by \Cref{eq:schedule} with $s=4$ and $r=0.25$, meaning it was the RNN that was trained the longest.}
    \label{tab:qmc_sublattice_magnetizations}
\end{table*}

\begin{table*}[h!]
    \centering
        \begin{tabularx}{\textwidth}{X X X}
            \hline\hline
        
        Boundary Condition & Architecture and reference & $E/N$ \\
        System Size $L\times L$ & & \\
        \hline
        periodic & RBM~\cite{carleo_solving_2017,wu_variational_2024} & $-0.668(5)$ \\
         $6\times6$ & CNN~\cite{choo_two-dimensional_2019} & $-0.67882(1)$\\
         & RBM+Lanczos~\cite{chen_systematic_2022}  & $-0.678868(2)$ \\
         & \bf{This work (best RNN)} & $\mathbf{-0.67887(2)}$\\
         & \bf{This work (zero-variance)} & $\mathbf{-0.67887177(7)}$ \\
        
        \hline
        open & LSTM~\cite{roth_iterative_2020} & $-0.603417$\\
         $6\times6$ & \bf{This work (zero-variance)} & $\mathbf{-0.60351496(1)}$ \\
         & RNN~\cite{hibat-allah_supplementing_2024} & $-0.603516(1)$\\
         & \bf{This work (best RNN)} & $\mathbf{-0.603517(6)}$ \\
         
        \hline
        periodic & CNN~\cite{choo_two-dimensional_2019} & $-0.67135$\\
         $10\times10$ & RBM+Lanczos~\cite{chen_systematic_2022} &  $-0.671519(4)$\\
         & \bf{This work (best RNN)} & $\mathbf{-0.67151(2)}$ \\
         & VMC - CNN+MinSR~\cite{chen_empowering_2024} & $-0.67155260(3)$ \\
         & \bf{This work (zero-variance)} & $\mathbf{-0.6715950(5)}$ \\
        
        \hline
        open & PixelCNN~\cite{sharir_deep_2020} & $-0.628627(1)$\\
        $10\times10$  & RNN~\cite{hibat-allah_supplementing_2024} & $-0.628638(4)$\\
        & \bf{This work (zero-variance)} & $\mathbf{-0.62864114(5)}$\\ 
        & \bf{This work (best RNN)} & $\mathbf{-0.628656(9)}$\\ 
        \hline\hline
    \end{tabularx}
    \caption{Neural quantum state results for the SLAHM on lattices with $L=6$ and $L=10$. For simplicity, only the first NQS results  and NQS results which improved upon existing variational energies were included in this table. The results we present in this work are in bold. The zero-variance energies for the SLAHM with open boundary conditions are obtained by extrapolating the variational energies from simulations with $s=2.0,4.0$. For the SLAHM with periodic boundary conditions, we perform the extrapolation using results from simulations with $s=1.0,2.0,4.0$.}
    \label{tab:nqs_energies}
\end{table*}

\begin{table*}[h!]
    \centering
    \begin{tabularx}{\textwidth}{X X X}
    \hline\hline
        Method and reference & $E_\infty/N$ & $M_\infty$  \\
        \hline
        Spin-Wave Theory~\cite{anderson_approximate_1952,kubo_spin-wave_1952,keffer_ferromagnetic_1960} &  $-0.6705$ & 0.303 \\
        \hline
        Series expansions~\cite{singh_thermodynamic_1989} & $-0.6696(3)$ & 0.313 \\
        \hline
        Exact Diagonalization~\cite{tang_long-range_1989} & $-0.672(1)$ & 0.25 \\       
        \hline
        QMC - World-line~\cite{reger_monte_1988} & $-0.670(1)$ & 0.31 \\
        QMC - Green's function~\cite{trivedi_green-function_1989} & $-0.6692(2)$ & 0.31(2) \\
        QMC - Green's function~\cite{,carlson_ground-state_1989} & $-0.6692(2)$ & 0.34(1) \\
        QMC - Stochastic Series Expansion~\cite{sandvik_finite-size_1997} & $-0.669437(5)$ & 0.3070(3) \\
        QMC - Stochastic Series Expansion~\cite{sandvik_loop_2010} & -- & 0.30743(1) \\
        \hline
        VMC - RVB ansatz~\cite{liang_new_1988} & $-0.6682(4)$ & -- \\
        VMC - Gutzwiller projector ansatz~\cite{yokoyama_hubbard_1987} & $-0.642$ & $\sim$0.43 \\
        VMC - Jastrow ansatz~\cite{huse_simple_1988} & $-0.6638$ & $\sim$0.4 \\
        VMC - Marshall-Jastrow ansatz~\cite{liu_variational_1989} & $-0.6637(2)$ & 0.349(2) \\
        \bf{VMC - RNN (this work)} & $\mathbf{-0.6694886(5)}$ & \bf{0.309(2), 0.308(2)} \\
        \hline
        two-dimensional DMRG~\cite{xiang_two-dimensional_2001} & $-0.6692$ & --\\
        DMRG (cylinders)~\cite{white_neel_2007} & -- & 0.3067\\
        gradient optimized iPEPS~\cite{hasik_investigation_2021} & $-0.669432$ & 0.3152 \\
        gradient optimized PEPS~\cite{liu_gradient_2017} & $-0.66948(42)$ & 0.305\\
        \hline\hline
    \end{tabularx}
    \caption{Values for the energy per spin and the sublattice magnetization of the SLAHM in the thermodynamic limit, adapted from~\cite{manousakis_spin-_1991}. This list is not comprehensive, but it summarizes the results discussed in the introduction and includes the state-of-the-art. Our results, which are shown in bold in this table, are those obtained from our simulations performed on lattices with periodic boundary conditions (e.g. the results presented in the main text).}
    \label{tab:TL}
\end{table*}

\clearpage
\twocolumngrid
\bibliography{main}

\end{document}